\documentclass[preprints,article,accept,moreauthors,pdftex]{Definitions/mdpi}
\usepackage{graphicx}

\firstpage{1} 
\pubvolume{1}
\issuenum{1}
\articlenumber{0}
\pubyear{2022}
\copyrightyear{2022}
\datereceived{} 
\dateaccepted{12 January 2023} 
\datepublished{} 
\hreflink{https://doi.org/} % If needed use \linebreak
\doinum{}
\Title{Multi-Wavelength and Multi-Messenger Studies with the next-generation Event Horizon Telescope}

\TitleCitation{Multi-wavelength and multi-messenger studies with the ngEHT.}

\Author{Rocco Lico $^{1,2,}$*\orcidA{}, Svetlana G.~Jorstad$^{3,4}$*\orcidB{}, Alan P.~Marscher $^{3}$\orcidC{}, Jose L.~G\'omez$^{1}$\orcidD{}, Ioannis Liodakis$^{5}$\orcidE{}, Rohan Dahale$^{1}$\orcidF{}, Antxon Alberdi$^{1}$\orcidG{}, Roman Gold$^{6}$\orcidH{}, Efthalia Traianou$^{1}$\orcidI{},Teresa Toscano$^{1}$\orcidJ{}, Marianna Foschi$^{1}$\orcidK{}.}

\AuthorCitation{Lico, R.; Jorstad, S.~G..; Marscher, A.~P., et al.}

\address{%
$^{1}$ \quad Instituto de Astrof\'{\i}sica de Andaluc\'{\i}a-CSIC, Glorieta de la Astronom\'{\i}a s/n, 18008 Granada, Spain.\\
$^{2}$ \quad INAF Istituto di Radioastronomia, via Gobetti 101, 40129 Bologna, Italy. \\
$^{3}$ \quad Institute for Astrophysical Research, Boston University, 725 Commonwealth Avenue, Boston, MA 02215. \\
$^{4}$ \quad Astronomical Institute, St. Petersburg State University, Universitetskij Pr. 28, Petrodvorets, St. Petersburg 198504, Russia. \\
$^{5}$ \quad Finnish Centre for Astronomy with ESO, FI-20014 University of Turku, Finland. \\ 
$^{6}$ \quad CP3-Origins, University of Southern Denmark, Campusvej 55, DK-5230 Odense M, Denmark. 
} 

\corres{Correspondence: rlico@iaa.es; jorstad@bu.edu.} 

\abstract{
The next-generation Event Horizon Telescope (ngEHT) will provide us with the best opportunity to investigate supermassive black holes (SMBHs) at the highest possible resolution and sensitivity. With respect to the existing Event Horizon Telescope (EHT) array, the ngEHT will provide increased sensitivity and uv-coverage with the addition of new stations, wider frequency coverage (from 86 GHz to 345 GHz and higher), finer resolution ($<15$ micro-arcseconds), and better monitoring capabilities.
This will offer a unique opportunity to deeply investigate the physics around SMBHs, such as the disk-jet connection, the mechanisms responsible for high-energy photon and neutrino events, the role of magnetic fields in shaping relativistic jets, as well as the nature of binary SMBH systems. In this white paper we describe some ngEHT science cases in the context of multi-wavelength studies and synergies.
}


\begin{document}

%%%%%%%%%%%%%%%%%%%%%%%%%%%%%%%%%%%%%%%%%%
%\setcounter{section}{-1} %% Remove this when starting to work on the template.

\section{Introduction}

Cosmic systems powered by accretion onto supermassive black holes (SMBHs) are the most intense sources of energy in the universe. The tremendous power they generate manifests in a variety of forms: electromagnetic waves from radio to $\gamma$-ray wavelengths, ultra-energetic particles, strong magnetic fields, and plasma jets propelled outwards at near-light speeds. There is now firm evidence that SMBHs lie at the hearts of nearly all galaxies. The first direct measurements of the sizes and masses of SMBHs have been obtained with the Event Horizon Telescope (EHT) at 230 GHz for the radio galaxy M87 (M$_{\texttt BH}=6.5\pm0.7\times$10$^9$M$_{Sun}$, \cite{EHTM87}) and the Milky Way black hole, SgrA* (M$_{\texttt BH}\approx$4$\times$10$^6$M$_{Sun}$, \cite{EHTSgrA}), which show consistency with the predictions of general relativity spanning over three orders of magnitude in central mass. Systems where matter falls into the black hole at a high rate create the phenomenon of active galactic nuclei (AGNs). The masses of black holes in AGNs range from less than 1 million to roughly 10 billion solar masses \citep[e.g., ][]{Woo2002,Ghez2003}. The luminosity of the jets can be as high as $\sim10^{15}$ times that of the sun. 

AGNs with powerful jets are rare in the universe, so most lie at distances of hundreds of millions or billions of light-years from the Earth. Much of their light comes from regions of similar size as our solar system. This extremely compact nature, combined with the great distances involved, causes these regions to appear extremely small on the sky. Observations designed to study AGNs therefore require telescopes with extraordinary resolving power. Furthermore, the extremely luminous emission, and thus the physical conditions that create it, are highly variable in time. Because of these two factors, instruments that probe the nature of AGNs and their jets must provide ultra-fine resolution imaging, as well as the ability to monitor rapid changes in the images. 

Such opportunities and capabilities will be offered in the near future by the next-generation Event Horizon Telescope (ngEHT). The main ngEHT goal is to upgrade and enhance the current capabilities of the EHT array by adding more than 10 new stations, improving the image dynamic range by $\sim2$ orders of magnitude, enabling simultaneous multi-frequency observations up to 345\,GHz and possibly higher frequencies, including polarization capabilities at all frequencies, and quadrupling the observing bandwidth \citep{Doeleman2019}.  
The ngEHT will have an angular resolution of $\sim10-15$ micro-arcseconds ($\mu as$), reaching linear scales of a fraction of a light-year. Moreover, the development of new super-resolution algorithms \citep[e.g.\,][]{Akiyama2017, Chael2018, Broderick2020}, will allow the smallest features resolved by the ngEHT to be further improved by a factor up to $\sim3$. Observations at 86 GHz will also be included, which are expected to be important for understanding the connection between the photon ring and inner jet. The ngEHT project development and implementation will consist of two phases. The first phase is based on the instrument design and site selection, and is expected to be concluded by the end of 2023. The second phase will consist of the actual construction and commissioning of the ngEHT array, expected to be fully operational by 2030, with gradual deployment over 2026-2030.

%%%%%%%%%%%%%%%%%%%%%%%%%%%%%%%%%%%%%%%%%%

\begin{figure}
\centering
\includegraphics[width=13.8 cm, trim={0 10.5cm 0 0}, clip]{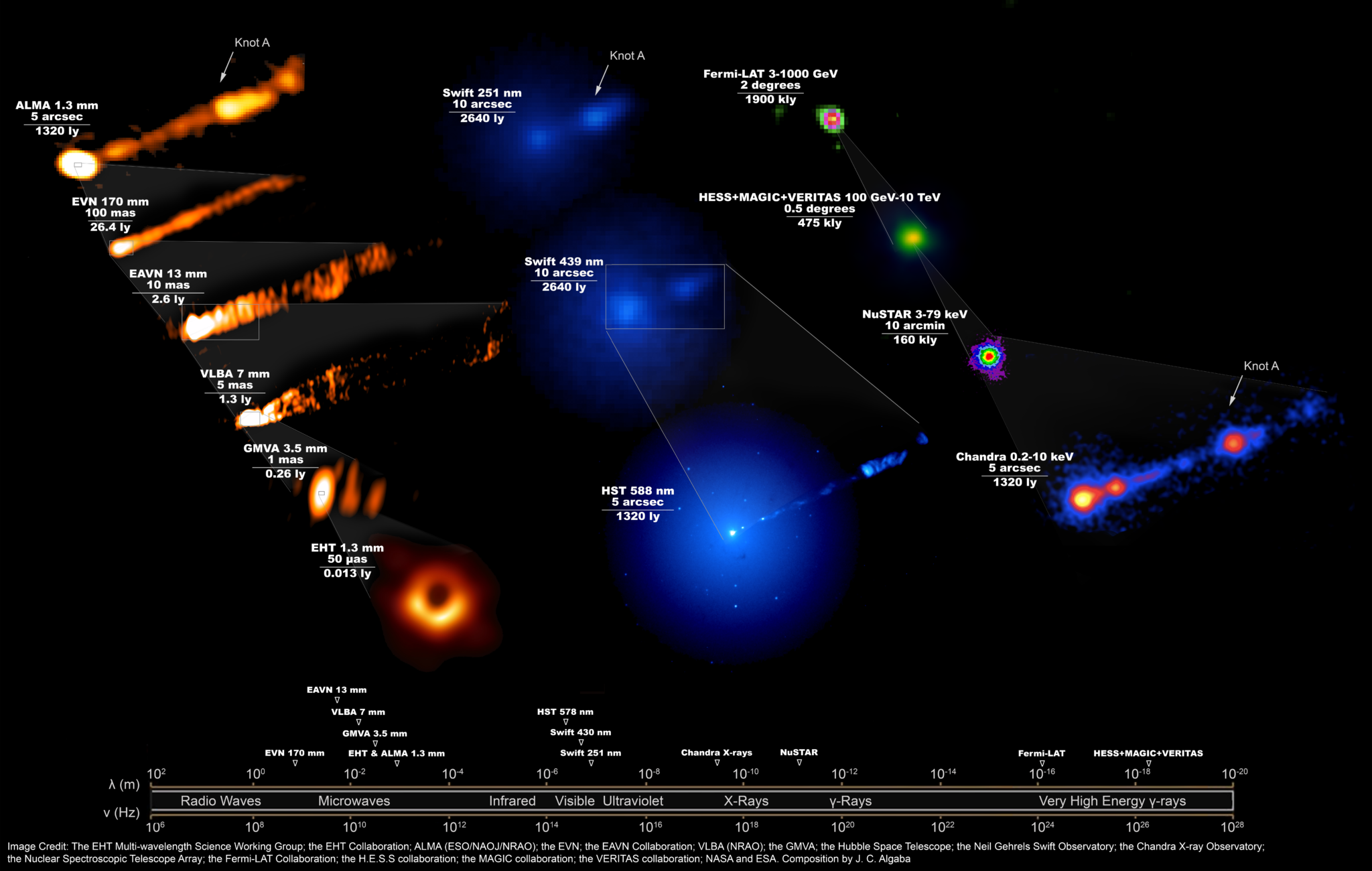}
\caption{Composite image of M87 as seen across the entire electromagnetic spectrum during the 2017 EHT observing campaign, adapted from \citep{EHT2021} under the terms of the CC BY 3.0 licence. For each image, the instrument, the observing wavelength, and the scale are shown next to it. The frequency-wavelength scale at the bottom indicates the location of each instrument in the electromagnetic spectrum. Image credit: The EHT Multi-wavelength Science Working Group; the EHT Collaboration; ALMA (ESO/NAOJ/NRAO); the EVN; the EAVN Collaboration; VLBA (NRAO); the GMVA; the Hubble Space Telescope; the Neil Gehrels Swift Observatory; the Chandra X-ray Observatory; the Nuclear Spectroscopic Telescope Array; the Fermi-LAT Collaboration; the MAGIC collaboration; the VERITAS collaboration; NASA an ESA. \label{eht_mwl}}
\end{figure}   
\unskip

\section{Science cases}
The EHT has already demonstrated the power of multi-wavelength (MWL) synergies for characterizing the underlying physics and emission mechanisms in AGNs, by means of an intensive multi-wavelength observing campaign of M87 conducted in 2017 with 19 different observing facilities (Fig.~\ref{eht_mwl}, \citet{EHT2021}). In this section, we describe some potential science cases for the ngEHT in the context of MWL studies.

\subsection{Disk-Jet connection}
The ngEHT will allow us to determine the jet's profile starting at its origin in the vicinity of a SMBH, the location and its effect on the jet and length of the jet flow's acceleration and collimation zone (ACZ), and the location of the Bondi radius (inside of which the SMBH's gravitational influence is important). This will provide crucial insights toward understanding the connection and interplay between the jet, accretion disk, and black hole. Currently, the jet-disk connection has only been reported for two radio galaxies, 3C~120 \cite{Marscher2002, Ritaban2009, Lohfink2013} and 3C~111 \cite{Ritaban2011, Tombesi2012}. Both these radio galaxies possess relativistic jets with apparent speeds of $\sim$5c \cite{Weaver2022}. They exhibit X-ray properties similar to Seyfert galaxies, which implies that the X-ray emission is produced in the accretion disk (plus its ``corona'' of hot electrons) in the vicinity of the SMBH. Monitoring with the Very Long Baseline Array (VLBA) at 43~GHz and X-ray satellites ({\it RXTE, Swift, and Suzaku}) established a connection between X-ray flux variations and emergence of new superluminal emission features (``knots'') in the jet: appearances of new knots are preceded by dips in the X-ray luminosity. This X-ray/radio connection conclusively demonstrates a link between changes in accretion disk structure and powerful ejection events. Moreover, there is a similarity with the behavior of X-ray binary systems (``microquasars''), allowing comparisons that are important for generalizing insights obtained regarding the origin of relativistic jets in both types of systems \cite{Fender2004}. 

The innermost jet profile and its evolution with distance from the SMBH are key factors to understand the jet collimation and acceleration. Recently, a number of studies have been devoted to this topic. Kovalev et al. \cite{Kovalev2020} have performed a search for transitions in the jet shape in a sample of radio-loud AGNs (367 sources) at 15 and 1.4~GHz. The authors have found that, for nearby AGNs (redshift $z<$0.07), 10 out of 29 jets exhibit a transition from a parabolic to a conical shape. They concluded that the transition in geometry is perhaps a common feature of AGNs jets, but can be observed only when sufficient linear resolution is achieved. The authors also suggested that the jet transition occurs when the particle kinetic energy becomes equal to the energy of the magnetic field, while the Bondi radius position is governed by ambient medium pressure and may not coincide with the transition break. Casadio et al. \cite{Casadio2021} have analyzed stacked VLBI images at 86 and 43 GHz of BL Lacertae (BL Lac, z=0.069) and found that its jet expands with a conical geometry, with a higher expansion rate between $\sim$5 and 10~pc (de-projected) from the BH. They proposed that the transition in the profile is connected with a change of the external pressure at the location of the Bondi radius ($\sim$3.3$\times$10$^5$R$_g$).  

The high-resolution ngEHT capabilities will allow direct imaging of these regions, tracing the jet down to the expected accretion disk and the areas where the jet is ejected and collimated in several AGNs. Such images and the studies they allow in combination with MWL observations, will provide decisive information for constructing and testing theoretical models.  

\begin{figure}%[H]
\centering
\includegraphics[width=9.5 cm]{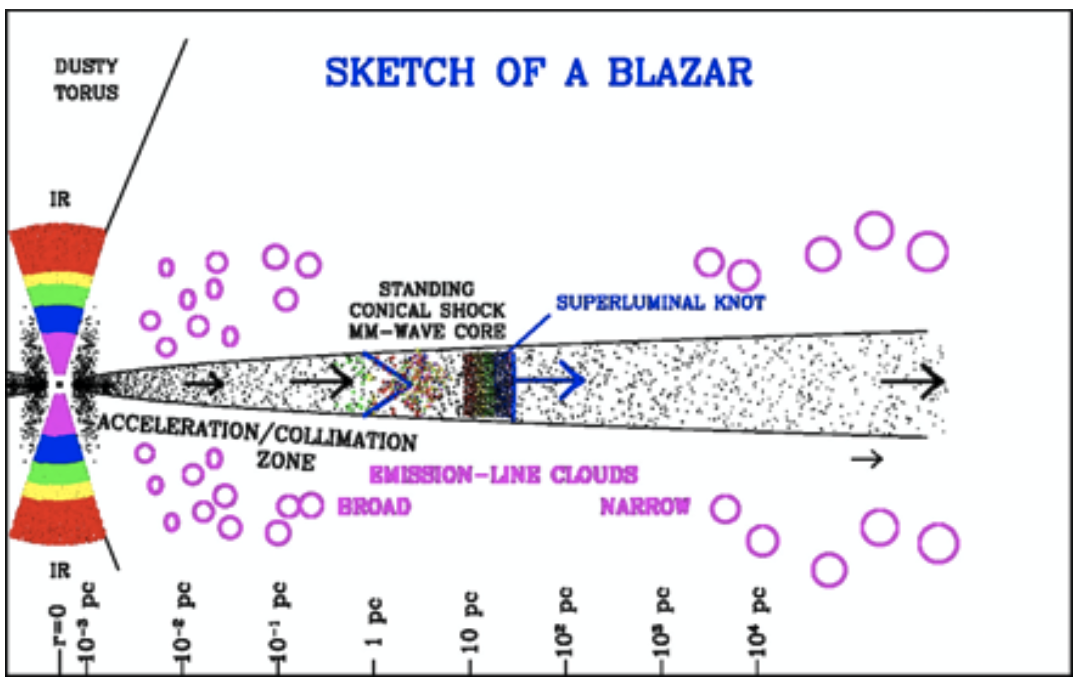}
\caption{Sketch of the innermost jet region in blazars (image credit: A.Marscher).\label{blazar_sketch}}
\end{figure}   
\unskip

\subsection{Nature of the VLBI core}
VLBI images of blazar jets, the most extreme objects in the family of AGNs, are dominated by a compact feature known as the "core". The nature of the core remains an open question, mostly because of optical depth effects and the limited angular resolution of VLBI arrays. At wavelengths longer than $\sim7$ mm, the core is optically thick and essentially unresolved, probably representing the area in the jet where the optical depth is $\sim1$. At 0.8-3 mm, where the resolution is higher and the core is optically thin, it should represent a physical structure in the jet's acceleration/collimation zone, such as a standing shock \citep{Liodakis2022}, kink in the jet flow, site of magnetic reconnection (Fig.~\ref{blazar_sketch}), or some other phenomenon. The EHT Collaboration has already imaged the core region of the blazar 3C\,279, finding it to be dynamic, with a puzzling multi-component structure \citep{kim2020}. The ngEHT will provide the data needed to make MWL movies to follow the motions, and perhaps the formation, of the components. This will lead to an understanding of the core, the brightest feature seen in relativistic jets, and how it connects to the SMBH and its accretion disk.     

A particularly important diagnostic of the nature of the core is its polarization. For example, a standing conical shock is expected to display a radial linear polarization pattern for viewing angles very close to the jet axis \citep{Cawthorne2013}, a pattern that can be distinguished by the ngEHT. A kink can develop from an instability of a helical field, which has the signature of a rotation measure gradient across the jet \citep[e.g.,][]{Gabuzda2015} that can be detected at the resolution of the ngEHT. Magnetic reconnection can occur where the jet's magnetic field is highly turbulent, with low polarization and chaotic position angles of polarization across the jet. They can also originate where the magnetic field lines form loops that are stretched parallel to the jet, in which case the linear polarization vectors should lie transversely to the jet.

\subsection{Spine-Sheath Structure}
High-frequency synchrotron peaked (HSP) blazars, such as Mrk~421 and Mrk~501, represent a long-standing problem: reconciling extreme luminosities of Tera-electronvolt (TeV) gamma rays, which require high relativistic Doppler beaming factors, with low apparent speeds observed in microwave VLBI images \citep[known as ``Doppler crisis''; e.g., ][]{Lico2012, Weaver2022}. One of the most favored models to resolve the Doppler crisis involves the presence of transverse structure in the jet, with an extremely fast (Lorentz factor $>20$) inner spine surrounded by a slower outer sheath \citep[e.g.\,][]{Ghisellini2005}. This would produce distinctive observable signatures in VLBI images, such as limb brightening, especially in linearly polarized intensity, with electric-vector position angles (EVPAs) that are transverse to the jet. All of these signatures have already been observed in the jets of both Mrk~421 and Mrk~501 \citep[e.g.\,][]{Giroletti2008, Lico2014, Koyama2019}, as well as in other blazars \citep[e.g.\,][]{Hovatta2012, Nagai2014, Gabuzda2015}. However, as it was previously mentioned, very-high angular resolution and sensitivity are needed to resolve the transverse structure in the innermost jet regions, and which will become routinely available with the ngEHT. In fact, for the case of Centaurus A, EHT observations at 228 GHz have revealed a highly collimated, asymmeytrically edge-brightened jet together with a fainter counterjet \cite{Janssen2021}. In addition, a spine-sheath structure was recently detected in the bright quasar 3C~273 with the fine resolution of space-VLBI {\it RadioAstron} observations at wavelengths of 18 and 6 cm \cite{Bruni2021}. At 18 cm, the jet exhibited limb-brightening that had not previously been seen in this source. At 6 cm, emission from the central stream of plasma was imaged, with a spatial distribution complementary to the limb-brightened emission, which indicates the presence of a spine. Regular monitoring with high resolution and sensitivity is needed to determine the jet stratification structure and its evolution. Such observations do not yet exist, but will be possible with the ngEHT, which will allow measurements of apparent speeds across the jets of several HSP blazars.

\begin{figure}%[H]
\centering
\includegraphics[width=9.5 cm]{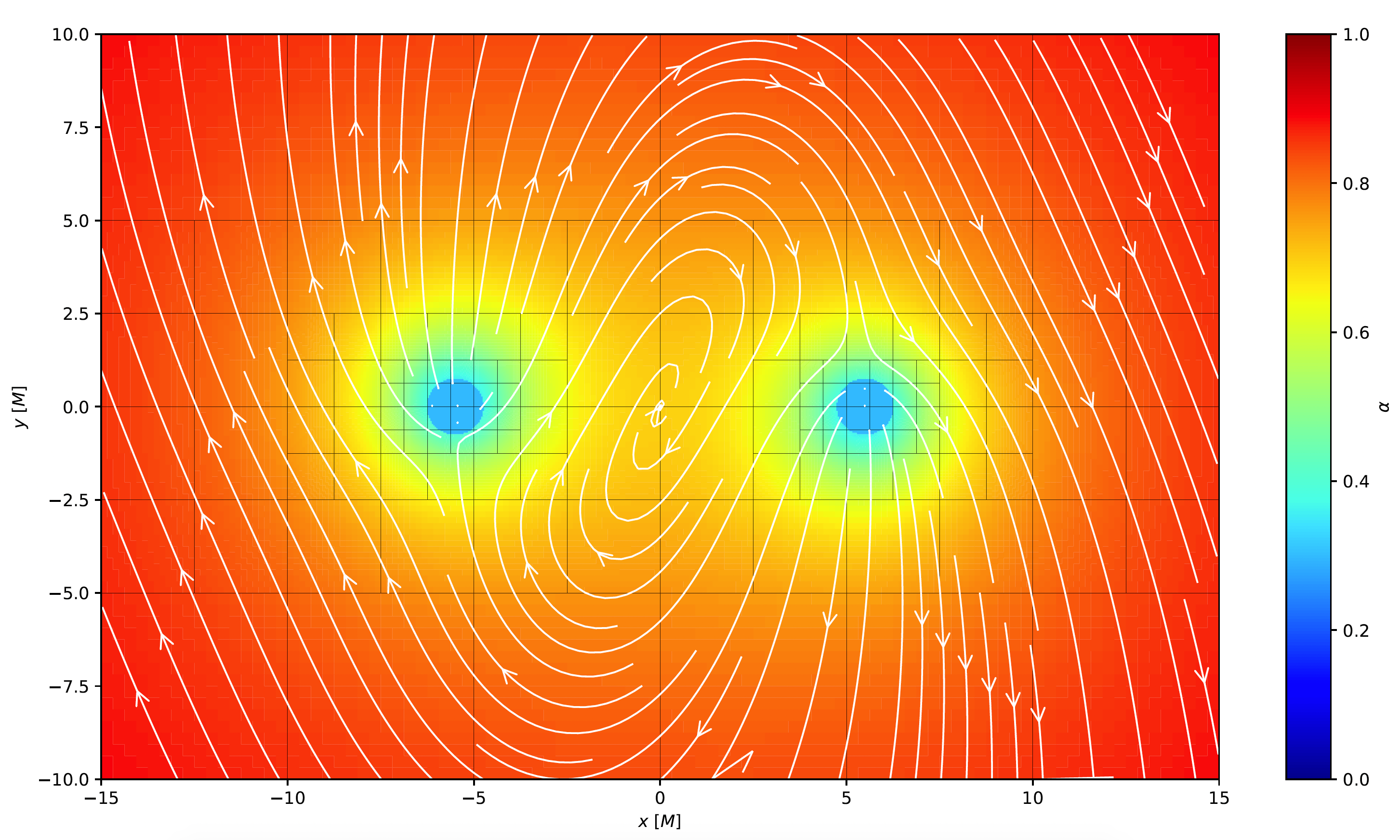}
\caption{Equatorial slice of a binary SMBH system in the corotating frame interpolated on the Black Hole Accretion Code (BHAC) adaptive mesh grid. The colorscale shows the spacetime lapse function and the arrows the shift vector.\label{binary_SMBH}}
\end{figure}   
\unskip

\subsection{Supermassive Binary Black Holes} \label{sect_smbh}
Binary SMBH systems represent a natural product of hierarchical galaxy formation and merger events. Currently, one of the best candidates for hosting a binary SMBH system is OJ\,287, a bright blazar showing quasi-periodic variability patterns in its optical light-curve that could be explained as the result of the secondary SMBH impacting the accretion disk of a primary one \cite{Valtonen2008, Gomez2022}. However, alternative scenarios to the binary SMBH system can also explain the variability pattern \citep[e.g., ][]{Liska2018}. 

Moreover, according to current theoretical models, when the system reaches an orbital separation of 0.1-10 parsec the dynamical processes that drive the coalescence of the two SMBHs cease to be effective and inhibit the merging process. This is known as ``the final parsec problem''. 
The ngEHT, with a resolution finer than $\sim15 \, \mu as$, will represent a unique opportunity to detect and spatially resolve from tens to hundreds sub-parsec binary SMBH systems (see \citet{Pesce2021} for more details about the statistics and the expected detections). Spatially resolving such binary SMBH systems would provide the ultimate proof of their existence. With its monitoring capabilities, the ngEHT will allow us to determine and characterize the orbital parameters of such tight binary systems, whose orbital periods are expected to range from months to a few years, and fully resolve their orbital motion. This is currently being investigated in set of numerical simulations by using the Black Hole Accretion Code \citep[BHAC, ][]{Porth2017}, see Fig.~\ref{binary_SMBH}.

Ongoing and upcoming large surveys in the optical to X-ray bands (e.g., the European-Chinese X-ray mission Einstein Probe) will provide prime lists of new binary SMBH candidates, among which the best-suited ones can be followed up with the ngEHT to check the source association and measure the orbital parameters. 
Moreover, the ngEHT high-resolution observations  would also represent an important resource for future gravitational wave detectors, both ground-based like the pulsar timing arrays (PTAs) and space-based like the Laser Interferometer Space Antenna (LISA), sensitive to the low-frequency band of the gravitational-wave spectrum expected from binary supermassive black holes.

\subsection{High-energy and neutrino events} \label{he}
Blazars are very efficient cosmic accelerators and have been proposed as the  possible source of several past high-energy neutrino events, although the production mechanisms and the precise region(s) where they form are still unknown \citep{Giommi2021}. The identification itself of the astrophysical counterparts of high-energy neutrinos is not straightforward (see e.g. \citep{Nanci2022}), and in this context the ngEHT could be helpful in two ways. On one hand, the high-resolution mm-wavelength ngEHT observations will allow us to identify and constrain the region(s) responsible for the neutrino production by tracing the variations and delay between the radio emission from the innermost jet regions and the neutrino detection event. On the other hand, owing to its monitoring and observing agility capabilities, the ngEHT will help to more accurately localize the neutrino emission sources by triggering prompt observations when a detection event occurs (see Y.~Y.~Kovalev et al., this issue of {\it Galaxies}).

The ngEHT will also provide essential details about the innermost regions of blazar jets, allowing us to disentangle the different regions where visible, X-ray, and $\gamma$-ray emission originates (e.g., near the SMBH, in the jet’s acceleration zone, in a standing shock, in a moving knot, in a kink, in regions of magnetic reconnection, etc.), and to better understand the mechanisms and physical processes responsible for the observed emission. The time-domain imaging capabilities of the ngEHT will allow matching of events in the jet with outbursts of high-energy photons with high accuracy. 
Moreover, multi-epoch and multi-frequency ngEHT polarization observations will allow investigations of the possible connection of magnetic field variations with the high-energy emission events \citep[e.g.,][]{Blinov2016}. This can be done by determining the time evolution of fundamental polarization parameters, such as the fractional polarization, rotation of the polarization plane, and Faraday rotation measure.

%In particular, the high-resolution mm-wavelength ngEHT observations will help to understand how and where protons, whose interactions with other particles are responsible for neutrino production, are accelerated to relativistic energies.

\subsection{Polarization and magnetic field studies}
The ngEHT will allow the characterization of the polarized radio emission in the near vicinity of the SMBH and constrain the magnetic field geometry, the plasma temperature, and the underlying particle density. This will determine whether the magnetic field is helical, as predicted by most theories, turbulent, or compressed by shocks \citep[e.g., ][]{Gomez2016, Issaoun2022, Zhao2022}. The presence of an ionized medium with a magnetic field causes the direction of linear polarization to rotate as radio waves pass through it; also known as Faraday rotation. MWL observations with the ngEHT can measure this effect to explore the medium both inside and surrounding the jet near its vertex. This can determine whether the jet is confined by a wind, an accretion flow, or its own magnetic field. 
Moreover, with the ngEHT it will be possible to measure the velocity field at the base of the jet, allowing us to test the different magneto-hydrodynamic jet formation models on scales of hundreds to thousands of Schwarzschild radii, and conclusively determine whether the jets are powered by the extraction of rotational energy from the spinning black hole \cite[Blandford-Znajek process, ][]{BZ1977} or from the accretion flow \cite[Blandford-Payne process, ][]{BP1982}.

Synergies with current (Imaging X-ray Polarimetry Explorer -- IXPE [\citealp{Weisskopf2022}]) and future (e.g., e-XTP [\citealp{Zhang2019}], COSI [\citealp{Beechert2022}]) high-energy polarimeters can also prove invaluable. The combination of broadband radio polarization from ngEHT and high-energy polarization observations will allow us to distinguish between shock and magnetic reconnection models for the particle energization in the jets \cite[e.g.,][]{Liodakis2022,DiGesu2022-Mrk421}, as well as accurately pin-point the particle acceleration regions. This could have significant implications for models of jet propagation as well as the nature of the blazar core (see section 2.2). It will also bring a unique view of the spatial evolution of the magnetic field configuration on diverse scales from the vicinity of the black hole to pc and kpc away.

\section{Technical requirements}

In order to measure Faraday rotation and gradients in the magnetic field of the most compact regions of jets, and other $\mu$as-scale emission regions, simultaneous multi-frequency ngEHT polarization observations are required (see A.~Ricarte et al., this issue of {\it Galaxies}). This can be implemented in several ways: (1) by using antennas whose frequency can be switched quickly, as is the case, for example, for the VLBA; (2) by combining ngEHT observations with other VLBI arrays (e.g., VLBA or next-generation VLA, at 43 and 86 GHz, respectively), or with other future space observatories (e.g., Millimetron); (3) by splitting the ngEHT into sub-arrays operating at different frequencies.

To achieve the science goals presented in this paper it will be essential to access the innermost jet regions, with typical 230 GHz flux densities ranging from a fraction to a few tens of mJy. For this reason, a sub-mJy sensitivity will be an essential requirement for the ngEHT, achievable thanks to the next-generation receivers and the addition of new telescopes to the array. More details about the sensitivity requirements will be presented in M.~Johnson et al.\ in prep.
In order to unambiguously connect MWL flaring and neutrino emission events to changes in the jets, monitoring capabilities are required for the ngEHT at a cadence shorter than the lifetime of the above-mentioned multi-messenger events. This could be possible during 1-2 week campaigns in the fortunate case that we are indeed observing a source with such an event. Either year-long or multiple weeks-long sessions per year would produce a high chance of success. Such multi-epoch ngEHT observations will need to be coordinated with other existing facilities, as well as with the new generation of instruments, such as the Cherenkov Telescope Array (CTA), LISA, and Athena. Moreover, the Russian-led next generation millimeter-band Space VLBI mission Millimetron, when launched, will provide a space arm to the ngEHT. This will enhance the resolution of the ngEHT to provide both space VLBI imaging, although with somewhat limited uv-coverage, and uv-domain studies of photon rings.

\authorcontributions{Conceptualization and writing---original draft preparation, R.L., S.G.J. and A.P.M.; writing---review and editing, all authors; All authors have read and agreed to the published version of the manuscript.}

\funding{RL, JLG and RD acknowledge financial support from the State Agency for Research of the Spanish MCIU through the “Center of Excellence Severo Ochoa” award for the Instituto de Astrofísica de Andalucía (SEV-2017-0709), from the Spanish Ministerio de Economía y Competitividad, and Ministerio de Ciencia e Innovaci\'on (grants AYA2016-80889-P, PID2019-108995GB-C21), the Consejería de Economía, Conocimiento, Empresas y Universidad of the Junta de Andalucía (grant P18-FR-1769), the Consejo Superior de Investigaciones Científicas (grant 2019AEP112). APM and SGJ acknowledge financial support from NASA Fermi Guest Investigator grants 80NSSC20K1567 and 80NSSC22K1571.}

\dataavailability{Not applicable.} 

%\acknowledgments{...}

\conflictsofinterest{The authors declare no conflict of interest.} 

%%%%%%%%%%%%%%%%%%%%%%%%%%%%%%%%%%%%%%%%%%
\begin{adjustwidth}{-\extralength}{0cm}

\reftitle{References}

\bibliography{main.bib}

\begin{thebibliography}{999}

\bibitem[{Event Horizon Telescope Collaboration} \em{et~al.}(2019){Event
  Horizon Telescope Collaboration}, {Akiyama}, {Alberdi}, {Alef}, {Asada},
  {Azulay}, {Baczko}, {Ball}, {Balokovi{\'c}}, {Barrett}, {Bintley},
  {Blackburn}, {Boland}, {Bouman}, {Bower}, {Bremer}, {Brinkerink},
  {Brissenden}, {Britzen}, {Broderick}, {Broguiere}, {Bronzwaer}, {Byun},
  {Carlstrom}, {Chael}, {Chan}, {Chatterjee}, {Chatterjee}, {Chen}, {Chen},
  {Cho}, {Christian}, {Conway}, {Cordes}, {Crew}, {Cui}, {Davelaar}, {De
  Laurentis}, {Deane}, {Dempsey}, {Desvignes}, {Dexter}, {Doeleman}, {Eatough},
  {Falcke}, {Fish}, {Fomalont}, {Fraga-Encinas}, {Freeman}, {Friberg}, {Fromm},
  {G{\'o}mez}, {Galison}, {Gammie}, {Garc{\'\i}a}, {Gentaz}, {Georgiev},
  {Goddi}, {Gold}, {Gu}, {Gurwell}, {Hada}, {Hecht}, {Hesper}, {Ho}, {Ho},
  {Honma}, {Huang}, {Huang}, {Hughes}, {Ikeda}, {Inoue}, {Issaoun}, {James},
  {Jannuzi}, {Janssen}, {Jeter}, {Jiang}, {Johnson}, {Jorstad}, {Jung},
  {Karami}, {Karuppusamy}, {Kawashima}, {Keating}, {Kettenis}, {Kim}, {Kim},
  {Kim}, {Kino}, {Koay}, {Koch}, {Koyama}, {Kramer}, {Kramer}, {Krichbaum},
  {Kuo}, {Lauer}, {Lee}, {Li}, {Li}, {Lindqvist}, {Liu}, {Liuzzo}, {Lo},
  {Lobanov}, {Loinard}, {Lonsdale}, {Lu}, {MacDonald}, {Mao}, {Markoff},
  {Marrone}, {Marscher}, {Mart{\'\i}-Vidal}, {Matsushita}, {Matthews},
  {Medeiros}, {Menten}, {Mizuno}, {Mizuno}, {Moran}, {Moriyama},
  {Moscibrodzka}, {M{\"u}ller}, {Nagai}, {Nagar}, {Nakamura}, {Narayan},
  {Narayanan}, {Natarajan}, {Neri}, {Ni}, {Noutsos}, {Okino}, {Olivares},
  {Ortiz-Le{\'o}n}, {Oyama}, {{\"O}zel}, {Palumbo}, {Patel}, {Pen}, {Pesce},
  {Pi{\'e}tu}, {Plambeck}, {PopStefanija}, {Porth}, {Prather},
  {Preciado-L{\'o}pez}, {Psaltis}, {Pu}, {Ramakrishnan}, {Rao}, {Rawlings},
  {Raymond}, {Rezzolla}, {Ripperda}, {Roelofs}, {Rogers}, {Ros}, {Rose},
  {Roshanineshat}, {Rottmann}, {Roy}, {Ruszczyk}, {Ryan}, {Rygl},
  {S{\'a}nchez}, {S{\'a}nchez-Arguelles}, {Sasada}, {Savolainen}, {Schloerb},
  {Schuster}, {Shao}, {Shen}, {Small}, {Sohn}, {SooHoo}, {Tazaki}, {Tiede},
  {Tilanus}, {Titus}, {Toma}, {Torne}, {Trent}, {Trippe}, {Tsuda}, {van
  Bemmel}, {van Langevelde}, {van Rossum}, {Wagner}, {Wardle}, {Weintroub},
  {Wex}, {Wharton}, {Wielgus}, {Wong}, {Wu}, {Young}, {Young}, {Younsi},
  {Yuan}, {Yuan}, {Zensus}, {Zhao}, {Zhao}, {Zhu}, {Algaba}, {Allardi},
  {Amestica}, {Anczarski}, {Bach}, {Baganoff}, {Beaudoin}, {Benson},
  {Berthold}, {Blanchard}, {Blundell}, {Bustamente}, {Cappallo},
  {Castillo-Dom{\'\i}nguez}, {Chang}, {Chang}, {Chang}, {Chen}, {Chilson},
  {Chuter}, {C{\'o}rdova Rosado}, {Coulson}, {Crawford}, {Crowley}, {David},
  {Derome}, {Dexter}, {Dornbusch}, {Dudevoir}, {Dzib}, {Eckart}, {Eckert},
  {Erickson}, {Everett}, {Faber}, {Farah}, {Fath}, {Folkers}, {Forbes},
  {Freund}, {G{\'o}mez-Ruiz}, {Gale}, {Gao}, {Geertsema}, {Graham}, {Greer},
  {Grosslein}, {Gueth}, {Haggard}, {Halverson}, {Han}, {Han}, {Hao},
  {Hasegawa}, {Henning}, {Hern{\'a}ndez-G{\'o}mez}, {Herrero-Illana},
  {Heyminck}, {Hirota}, {Hoge}, {Huang}, {Impellizzeri}, {Jiang}, {Kamble},
  {Keisler}, {Kimura}, {Kono}, {Kubo}, {Kuroda}, {Lacasse}, {Laing}, {Leitch},
  {Li}, {Lin}, {Liu}, {Liu}, {Lu}, {Marson}, {Martin-Cocher}, {Massingill},
  {Matulonis}, {McColl}, {McWhirter}, {Messias}, {Meyer-Zhao}, {Michalik},
  {Monta{\~n}a}, {Montgomerie}, {Mora-Klein}, {Muders}, {Nadolski}, {Navarro},
  {Neilsen}, {Nguyen}, {Nishioka}, {Norton}, {Nowak}, {Nystrom}, {Ogawa},
  {Oshiro}, {Oyama}, {Parsons}, {Paine}, {Pe{\~n}alver}, {Phillips}, {Poirier},
  {Pradel}, {Primiani}, {Raffin}, {Rahlin}, {Reiland}, {Risacher}, {Ruiz},
  {S{\'a}ez-Mada{\'\i}n}, {Sassella}, {Schellart}, {Shaw}, {Silva}, {Shiokawa},
  {Smith}, {Snow}, {Souccar}, {Sousa}, {Sridharan}, {Srinivasan}, {Stahm},
  {Stark}, {Story}, {Timmer}, {Vertatschitsch}, {Walther}, {Wei}, {Whitehorn},
  {Whitney}, {Woody}, {Wouterloot}, {Wright}, {Yamaguchi}, {Yu}, {Zeballos},
  {Zhang}, and {Ziurys}]{EHTM87}
{Event Horizon Telescope Collaboration}.; {Akiyama}, K.; {Alberdi}, A.; {Alef},
  W.; {Asada}, K.; {Azulay}, R.; {Baczko}, A.K.; {Ball}, D.; {Balokovi{\'c}},
  M.; {Barrett}, J.;  et~al.
\newblock {First M87 Event Horizon Telescope Results. I. The Shadow of the
  Supermassive Black Hole}.
\newblock {\em \apjl} {\bf 2019}, {\em 875},~L1,
  \href{http://xxx.lanl.gov/abs/1906.11238}{{\normalfont
  [arXiv:astro-ph.GA/1906.11238]}}.
\newblock {\url{https://doi.org/10.3847/2041-8213/ab0ec7}}.

\bibitem[{Event Horizon Telescope Collaboration} \em{et~al.}(2022){Event
  Horizon Telescope Collaboration}, {Akiyama}, {Alberdi}, {Alef}, {Algaba},
  {Anantua}, {Asada}, {Azulay}, {Bach}, {Baczko}, {Ball}, {Balokovi{\'c}},
  {Barrett}, {Baub{\"o}ck}, {Benson}, {Bintley}, {Blackburn}, {Blundell},
  {Bouman}, {Bower}, {Boyce}, {Bremer}, {Brinkerink}, {Brissenden}, {Britzen},
  {Broderick}, {Broguiere}, {Bronzwaer}, {Bustamante}, {Byun}, {Carlstrom},
  {Ceccobello}, {Chael}, {Chan}, {Chatterjee}, {Chatterjee}, {Chen}, {Chen},
  {Cheng}, {Cho}, {Christian}, {Conroy}, {Conway}, {Cordes}, {Crawford},
  {Crew}, {Cruz-Osorio}, {Cui}, {Davelaar}, {Laurentis}, {Deane}, {Dempsey},
  {Desvignes}, {Dexter}, {Dhruv}, {Doeleman}, {Dougal}, {Dzib}, {Eatough},
  {Emami}, {Falcke}, {Farah}, {Fish}, {Fomalont}, {Ford}, {Fraga-Encinas},
  {Freeman}, {Friberg}, {Fromm}, {Fuentes}, {Galison}, {Gammie}, {Garc{\'\i}a},
  {Gentaz}, {Georgiev}, {Goddi}, {Gold}, {G{\'o}mez-Ruiz}, {G{\'o}mez}, {Gu},
  {Gurwell}, {Hada}, {Haggard}, {Haworth}, {Hecht}, {Hesper}, {Heumann}, {Ho},
  {Ho}, {Honma}, {Huang}, {Huang}, {Hughes}, {Ikeda}, {Impellizzeri}, {Inoue},
  {Issaoun}, {James}, {Jannuzi}, {Janssen}, {Jeter}, {Jiang},
  {Jim{\'e}nez-Rosales}, {Johnson}, {Jorstad}, {Joshi}, {Jung}, {Karami},
  {Karuppusamy}, {Kawashima}, {Keating}, {Kettenis}, {Kim}, {Kim}, {Kim},
  {Kim}, {Kino}, {Koay}, {Kocherlakota}, {Kofuji}, {Koch}, {Koyama}, {Kramer},
  {Kramer}, {Krichbaum}, {Kuo}, {Bella}, {Lauer}, {Lee}, {Lee}, {Leung},
  {Levis}, {Li}, {Lico}, {Lindahl}, {Lindqvist}, {Lisakov}, {Liu}, {Liu},
  {Liuzzo}, {Lo}, {Lobanov}, {Loinard}, {Lonsdale}, {Lu}, {Mao}, {Marchili},
  {Markoff}, {Marrone}, {Marscher}, {Mart{\'\i}-Vidal}, {Matsushita},
  {Matthews}, {Medeiros}, {Menten}, {Michalik}, {Mizuno}, {Mizuno}, {Moran},
  {Moriyama}, {Moscibrodzka}, {M{\"u}ller}, {Mus}, {Musoke}, {Myserlis},
  {Nadolski}, {Nagai}, {Nagar}, {Nakamura}, {Narayan}, {Narayanan},
  {Natarajan}, {Nathanail}, {Fuentes}, {Neilsen}, {Neri}, {Ni}, {Noutsos},
  {Nowak}, {Oh}, {Okino}, {Olivares}, {Ortiz-Le{\'o}n}, {Oyama}, {{\"O}zel},
  {Palumbo}, {Paraschos}, {Park}, {Parsons}, {Patel}, {Pen}, {Pesce},
  {Pi{\'e}tu}, {Plambeck}, {PopStefanija}, {Porth}, {P{\"o}tzl}, {Prather},
  {Preciado-L{\'o}pez}, {Psaltis}, {Pu}, {Ramakrishnan}, {Rao}, {Rawlings},
  {Raymond}, {Rezzolla}, {Ricarte}, {Ripperda}, {Roelofs}, {Rogers}, {Ros},
  {Romero-Ca{\~n}izales}, {Roshanineshat}, {Rottmann}, {Roy}, {Ruiz},
  {Ruszczyk}, {Rygl}, {S{\'a}nchez}, {S{\'a}nchez-Arg{\"u}elles},
  {S{\'a}nchez-Portal}, {Sasada}, {Satapathy}, {Savolainen}, {Schloerb},
  {Schonfeld}, {Schuster}, {Shao}, {Shen}, {Small}, {Sohn}, {SooHoo},
  {Souccar}, {Sun}, {Tazaki}, {Tetarenko}, {Tiede}, {Tilanus}, {Titus},
  {Torne}, {Traianou}, {Trent}, {Trippe}, {Turk}, {van Bemmel}, {van
  Langevelde}, {van Rossum}, {Vos}, {Wagner}, {Ward-Thompson}, {Wardle},
  {Weintroub}, {Wex}, {Wharton}, {Wielgus}, {Wiik}, {Witzel}, {Wondrak},
  {Wong}, {Wu}, {Yamaguchi}, {Yoon}, {Young}, {Young}, {Younsi}, {Yuan},
  {Yuan}, {Zensus}, {Zhang}, {Zhao}, {Zhao}, {Agurto}, {Allardi}, {Amestica},
  {Araneda}, {Arriagada}, {Berghuis}, {Bertarini}, {Berthold}, {Blanchard},
  {Brown}, {C{\'a}rdenas}, {Cantzler}, {Caro}, {Castillo-Dom{\'\i}nguez},
  {Chan}, {Chang}, {Chang}, {Chang}, {Chang}, {Chen}, {Chilson}, {Chuter},
  {Ciechanowicz}, {Colin-Beltran}, {Coulson}, {Crowley}, {Degenaar},
  {Dornbusch}, {Dur{\'a}n}, {Everett}, {Faber}, {Forster}, {Fuchs}, {Gale},
  {Geertsema}, {Gonz{\'a}lez}, {Graham}, {Gueth}, {Halverson}, {Han}, {Han},
  {Hasegawa}, {Hern{\'a}ndez-Rebollar}, {Herrera}, {Herrero-Illana},
  {Heyminck}, {Hirota}, {Hoge}, {Hostler Schimpf}, {Howie}, {Huang}, {Jiang},
  {Jinchi}, {John}, {Kimura}, {Klein}, {Kubo}, {Kuroda}, {Kwon}, {Lacasse},
  {Laing}, {Leitch}, {Li}, {Liu}, {Liu}, {Lin}, {Lu}, {Mac-Auliffe},
  {Martin-Cocher}, {Matulonis}, {Maute}, {Messias}, {Meyer-Zhao},
  {Monta{\~n}a}, {Montenegro-Montes}, {Montgomerie}, {Moreno Nolasco},
  {Muders}, {Nishioka}, {Norton}, {Nystrom}, {Ogawa}, {Olivares}, {Oshiro},
  {P{\'e}rez-Beaupuits}, {Parra}, {Phillips}, {Poirier}, {Pradel}, {Qiu},
  {Raffin}, {Rahlin}, {Ram{\'\i}rez}, {Ressler}, {Reynolds},
  {Rodr{\'\i}guez-Montoya}, {Saez-Madain}, {Santana}, {Shaw}, {Shirkey},
  {Silva}, {Snow}, {Sousa}, {Sridharan}, {Stahm}, {Stark}, {Test},
  {Torstensson}, {Venegas}, {Walther}, {Wei}, {White}, {Wieching}, {Wijnands},
  {Wouterloot}, {Yu}, {Yu (于威)}, and {Zeballos}]{EHTSgrA}
{Event Horizon Telescope Collaboration}.; {Akiyama}, K.; {Alberdi}, A.; {Alef},
  W.; {Algaba}, J.C.; {Anantua}, R.; {Asada}, K.; {Azulay}, R.; {Bach}, U.;
  {Baczko}, A.K.;  et~al.
\newblock {First Sagittarius A* Event Horizon Telescope Results. I. The Shadow
  of the Supermassive Black Hole in the Center of the Milky Way}.
\newblock {\em \apjl} {\bf 2022}, {\em 930},~L12.
\newblock {\url{https://doi.org/10.3847/2041-8213/ac6674}}.

\bibitem[{Woo} and {Urry}(2002)]{Woo2002}
{Woo}, J.H.; {Urry}, C.M.
\newblock {Active Galactic Nucleus Black Hole Masses and Bolometric
  Luminosities}.
\newblock {\em \apj} {\bf 2002}, {\em 579},~530--544,
  \href{http://xxx.lanl.gov/abs/astro-ph/0207249}{{\normalfont
  [arXiv:astro-ph/astro-ph/0207249]}}.
\newblock {\url{https://doi.org/10.1086/342878}}.

\bibitem[{Ghez} \em{et~al.}(2003){Ghez}, {Duch{\^e}ne}, {Matthews},
  {Hornstein}, {Tanner}, {Larkin}, {Morris}, {Becklin}, {Salim}, {Kremenek},
  {Thompson}, {Soifer}, {Neugebauer}, and {McLean}]{Ghez2003}
{Ghez}, A.M.; {Duch{\^e}ne}, G.; {Matthews}, K.; {Hornstein}, S.D.; {Tanner},
  A.; {Larkin}, J.; {Morris}, M.; {Becklin}, E.E.; {Salim}, S.; {Kremenek}, T.;
   et~al.
\newblock {The First Measurement of Spectral Lines in a Short-Period Star Bound
  to the Galaxy's Central Black Hole: A Paradox of Youth}.
\newblock {\em \apjl} {\bf 2003}, {\em 586},~L127--L131,
  \href{http://xxx.lanl.gov/abs/astro-ph/0302299}{{\normalfont
  [arXiv:astro-ph/astro-ph/0302299]}}.
\newblock {\url{https://doi.org/10.1086/374804}}.

\bibitem[{Doeleman} \em{et~al.}(2019){Doeleman}, {Blackburn}, {Dexter},
  {Gomez}, {Johnson}, {Palumbo}, {Weintroub}, {Farah}, {Fish}, {Loinard},
  {Lonsdale}, {Narayanan}, {Patel}, {Pesce}, {Raymond}, {Tilanus}, {Wielgus},
  {Akiyama}, {Bower}, {Broderick}, {Deane}, {Fromm}, {Gammie}, {Gold},
  {Janssen}, {Kawashima}, {Krichbaum}, {Marrone}, {Matthews}, {Mizuno},
  {Rezzolla}, {Roelofs}, {Ros}, {Savolainen}, {Yuan}, {Zhao}, {Blackburn},
  {Doeleman}, {Dexter}, {Gomez}, {Johnson}, {Palumbo}, {Weintroub}, {Farah},
  {Fish}, {Loinard}, {Lonsdale}, {Narayanan}, {Patel}, {Pesce}, {Raymond},
  {Tilanus}, {Wielgus}, {Akiyama}, {Bower}, {Broderick}, {Deane}, {Fromm},
  {Gammie}, {Gold}, {Janssen}, {Kawashima}, {Krichbaum}, {Marrone}, {Matthews},
  {Mizuno}, {Rezzolla}, {Roelofs}, {Ros}, {Savolainen}, {Yuan}, and
  {Zhao}]{Doeleman2019}
{Doeleman}, S.; {Blackburn}, L.; {Dexter}, J.; {Gomez}, J.L.; {Johnson}, M.D.;
  {Palumbo}, D.C.; {Weintroub}, J.; {Farah}, J.R.; {Fish}, V.; {Loinard}, L.;
  et~al.
\newblock {Studying Black Holes on Horizon Scales with VLBI Ground Arrays}.
\newblock In Proceedings of the Bulletin of the American Astronomical Society,
  2019, Vol.~51, p. 256,
  \href{http://xxx.lanl.gov/abs/1909.01411}{{\normalfont
  [arXiv:astro-ph.IM/1909.01411]}}.

\bibitem[{Akiyama} \em{et~al.}(2017){Akiyama}, {Kuramochi}, {Ikeda}, {Fish},
  {Tazaki}, {Honma}, {Doeleman}, {Broderick}, {Dexter}, {Mo{\'s}cibrodzka},
  {Bouman}, {Chael}, and {Zaizen}]{Akiyama2017}
{Akiyama}, K.; {Kuramochi}, K.; {Ikeda}, S.; {Fish}, V.L.; {Tazaki}, F.;
  {Honma}, M.; {Doeleman}, S.S.; {Broderick}, A.E.; {Dexter}, J.;
  {Mo{\'s}cibrodzka}, M.;  et~al.
\newblock {Imaging the Schwarzschild-radius-scale Structure of M87 with the
  Event Horizon Telescope Using Sparse Modeling}.
\newblock {\em \apj} {\bf 2017}, {\em 838},~1,
  \href{http://xxx.lanl.gov/abs/1702.07361}{{\normalfont
  [arXiv:astro-ph.IM/1702.07361]}}.
\newblock {\url{https://doi.org/10.3847/1538-4357/aa6305}}.

\bibitem[{Chael} \em{et~al.}(2018){Chael}, {Johnson}, {Bouman}, {Blackburn},
  {Akiyama}, and {Narayan}]{Chael2018}
{Chael}, A.A.; {Johnson}, M.D.; {Bouman}, K.L.; {Blackburn}, L.L.; {Akiyama},
  K.; {Narayan}, R.
\newblock {Interferometric Imaging Directly with Closure Phases and Closure
  Amplitudes}.
\newblock {\em \apj} {\bf 2018}, {\em 857},~23,
  \href{http://xxx.lanl.gov/abs/1803.07088}{{\normalfont
  [arXiv:astro-ph.IM/1803.07088]}}.
\newblock {\url{https://doi.org/10.3847/1538-4357/aab6a8}}.

\bibitem[{Broderick} \em{et~al.}(2020){Broderick}, {Pesce}, {Tiede}, {Pu}, and
  {Gold}]{Broderick2020}
{Broderick}, A.E.; {Pesce}, D.W.; {Tiede}, P.; {Pu}, H.Y.; {Gold}, R.
\newblock {Hybrid Very Long Baseline Interferometry Imaging and Modeling with
  THEMIS}.
\newblock {\em \apj} {\bf 2020}, {\em 898},~9,
  \href{http://xxx.lanl.gov/abs/2208.09003}{{\normalfont
  [arXiv:astro-ph.IM/2208.09003]}}.
\newblock {\url{https://doi.org/10.3847/1538-4357/ab9c1f}}.

\bibitem[{EHT MWL Science Working Group} \em{et~al.}(2021){EHT MWL Science
  Working Group}, {Algaba}, {Anczarski}, {Asada}, {Balokovi{\'c}}, {Chandra},
  {Cui}, {Falcone}, {Giroletti}, {Goddi}, {Hada}, {Haggard}, {Jorstad}, {Kaur},
  {Kawashima}, {Keating}, {Kim}, {Kino}, {Komossa}, {Kravchenko}, {Krichbaum},
  {Lee}, {Lu}, {Lucchini}, {Markoff}, {Neilsen}, {Nowak}, {Park}, {Principe},
  {Ramakrishnan}, {Reynolds}, {Sasada}, {Savchenko}, {Williamson}, {Event
  Horizon Telescope Collaboration}, {Akiyama}, {Alberdi}, {Alef}, {Anantua},
  {Azulay}, {Baczko}, {Ball}, {Barrett}, {Bintley}, {Benson}, {Blackburn},
  {Blundell}, {Boland}, {Bouman}, {Bower}, {Boyce}, {Bremer}, {Brinkerink},
  {Brissenden}, {Britzen}, {Broderick}, {Broguiere}, {Bronzwaer}, {Byun},
  {Carlstrom}, {Chael}, {Chan}, {Chatterjee}, {Chatterjee}, {Chen}, {Chen},
  {Chesler}, {Cho}, {Christian}, {Conway}, {Cordes}, {Crawford}, {Crew},
  {Cruz-Osorio}, {Davelaar}, {de Laurentis}, {Deane}, {Dempsey}, {Desvignes},
  {Dexter}, {Doeleman}, {Eatough}, {Falcke}, {Farah}, {Fish}, {Fomalont},
  {Ford}, {Fraga-Encinas}, {Friberg}, {Fromm}, {Fuentes}, {Galison}, {Gammie},
  {Garc{\'\i}a}, {Gentaz}, {Georgiev}, {Gold}, {G{\'o}mez}, {G{\'o}mez-Ruiz},
  {Gu}, {Gurwell}, {Hecht}, {Hesper}, {Ho}, {Ho}, {Honma}, {Huang}, {Huang},
  {Hughes}, {Ikeda}, {Inoue}, {Issaoun}, {James}, {Jannuzi}, {Janssen},
  {Jeter}, {Jiang}, {Jim{\'e}nez-Rosales}, {Johnson}, {Jung}, {Karami},
  {Karuppusamy}, {Kettenis}, {Kim}, {Kim}, {Kim}, {Koay}, {Kofuji}, {Koch},
  {Koyama}, {Kramer}, {Kramer}, {Kuo}, {Lauer}, {Levis}, {Li}, {Li},
  {Lindqvist}, {Lico}, {Lindahl}, {Liu}, {Liu}, {Liuzzo}, {Lo}, {Lobanov},
  {Loinard}, {Lonsdale}, {MacDonald}, {Mao}, {Marchili}, {Marrone}, {Marscher},
  {Mart{\'\i}-Vidal}, {Matsushita}, {Matthews}, {Medeiros}, {Menten}, {Mizuno},
  {Mizuno}, {Moran}, {Moriyama}, {Moscibrodzka}, {M{\"u}ller}, {Musoke},
  {Mej{\'\i}as}, {Nagai}, {Nagar}, {Nakamura}, {Narayan}, {Narayanan},
  {Natarajan}, {Nathanail}, {Neri}, {Ni}, {Noutsos}, {Okino}, {Olivares},
  {Ortiz-Le{\'o}n}, {Oyama}, {{\"O}zel}, {Palumbo}, {Patel}, {Pen}, {Pesce},
  {Pi{\'e}tu}, {Plambeck}, {Popstefanija}, {Porth}, {P{\"o}tzl}, {Prather},
  {Preciado-L{\'o}pez}, {Psaltis}, {Pu}, {Rao}, {Rawlings}, {Raymond},
  {Rezzolla}, {Ricarte}, {Ripperda}, {Roelofs}, {Rogers}, {Ros}, {Rose},
  {Roshanineshat}, {Rottmann}, {Roy}, {Ruszczyk}, {Rygl}, {S{\'a}nchez},
  {S{\'a}nchez-Arguelles}, {Savolainen}, {Schloerb}, {Schuster}, {Shao},
  {Shen}, {Small}, {Sohn}, {Soohoo}, {Sun}, {Tazaki}, {Tetarenko}, {Tiede},
  {Tilanus}, {Titus}, {Toma}, {Torne}, {Trent}, {Traianou}, {Trippe}, {van
  Bemmel}, {van Langevelde}, {van Rossum}, {Wagner}, {Ward-Thompson}, {Wardle},
  {Weintroub}, {Wex}, {Wharton}, {Wielgus}, {Wong}, {Wu}, {Yoon}, {Young},
  {Young}, {Younsi}, {Yuan}, {Yuan}, {Zensus}, {Zhao}, {Zhao}, {Fermi Large
  Area Telescope Collaboration}, {Principe}, {Giroletti}, {D'Ammando},
  {Orienti}, {H.~E.~S.~S. Collaboration}, {Abdalla}, {Adam}, {Aharonian},
  {Benkhali}, {Ang{\"u}ner}, {Arcaro}, {Armand}, {Armstrong}, {Ashkar},
  {Backes}, {Baghmanyan}, {Barbosa Martins}, {Barnacka}, {Barnard},
  {Becherini}, {Berge}, {Bernl{\"o}hr}, {Bi}, {B{\"o}ttcher}, {Boisson},
  {Bolmont}, {de Lavergne}, {Breuhaus}, {Brun}, {Brun}, {Bryan}, {B{\"u}chele},
  {Bulik}, {Bylund}, {Caroff}, {Carosi}, {Casanova}, {Chand}, {Chen}, {Cotter},
  {Cury{\l}o}, {Damascene Mbarubucyeye}, {Davids}, {Davies}, {Deil}, {Devin},
  {Dewilt}, {Dirson}, {Djannati-Ata{\"\i}}, {Dmytriiev}, {Donath},
  {Doroshenko}, {Duffy}, {Dyks}, {Egberts}, {Eichhorn}, {Einecke}, {Emery},
  {Ernenwein}, {Feijen}, {Fegan}, {Fiasson}, {de Clairfontaine}, {Fontaine},
  {Funk}, {F{\"u}{\ss}ling}, {Gabici}, {Gallant}, {Giavitto}, {Giunti},
  {Glawion}, {Glicenstein}, {Gottschall}, {Grondin}, {Hahn}, {Haupt},
  {Hermann}, {Hinton}, {Hofmann}, {Hoischen}, {Holch}, {Holler}, {H{\"o}rbe},
  {Horns}, {Huber}, {Jamrozy}, {Jankowsky}, {Jankowsky}, {Jardin-Blicq},
  {Joshi}, {Jung-Richardt}, {Kasai}, {Kastendieck}, {Katarzy{\'n}ski}, {Katz},
  {Khangulyan}, {Kh{\'e}lifi}, {Klepser}, {Klu{\'z}niak}, {Komin}, {Konno},
  {Kosack}, {Kostunin}, {Kreter}, {Lamanna}, {Lemi{\`e}re}, {Lemoine-Goumard},
  {Lenain}, {Levy}, {Lohse}, {Lypova}, {Mackey}, {Majumdar}, {Malyshev},
  {Malyshev}, {Marandon}, {Marchegiani}, {Marcowith}, {Mares},
  {Mart{\'\i}-Devesa}, {Marx}, {Maurin}, {Meintjes}, {Meyer}, {Moderski},
  {Mohamed}, {Mohrmann}, {Montanari}, {Moore}, {Morris}, {Moulin}, {Muller},
  {Murach}, {Nakashima}, {Nayerhoda}, {de Naurois}, {Ndiyavala},
  {Niederwanger}, {Niemiec}, {Oakes}, {O'Brien}, {Odaka}, {Ohm},
  {Olivera-Nieto}, {de Ona Wilhelmi}, {Ostrowski}, {Panter}, {Panny},
  {Parsons}, {Peron}, {Peyaud}, {Piel}, {Pita}, {Poireau}, {Noel}, {Prokhorov},
  {Prokoph}, {P{\"u}hlhofer}, {Punch}, {Quirrenbach}, {Rauth}, {Reichherzer},
  {Reimer}, {Reimer}, {Remy}, {Renaud}, {Rieger}, {Rinchiuso}, {Romoli},
  {Rowell}, {Rudak}, {Ruiz-Velasco}, {Sahakian}, {Sailer}, {Sanchez},
  {Santangelo}, {Sasaki}, {Scalici}, {Schutte}, {Schwanke}, {Schwemmer},
  {Seglar-Arroyo}, {Senniappan}, {Seyffert}, {Shafi}, {Shiningayamwe},
  {Simoni}, {Sinha}, {Sol}, {Specovius}, {Spencer}, {Spir-Jacob}, {Stawarz},
  {Sun}, {Steenkamp}, {Stegmann}, {Steinmassl}, {Steppa}, {Takahashi},
  {Tavernier}, {Taylor}, {Terrier}, {Tiziani}, {Tluczykont}, {Tomankova},
  {Trichard}, {Tsirou}, {Tuffs}, {Uchiyama}, {van der Walt}, {van Eldik}, {van
  Rensburg}, {van Soelen}, {Vasileiadis}, {Veh}, {Venter}, {Vincent}, {Vink},
  {V{\"o}lk}, {Vuillaume}, {Wadiasingh}, {Wagner}, {Watson}, {Werner}, {White},
  {Wierzcholska}, {Wong}, {Yusafzai}, {Zacharias}, {Zanin}, {Zargaryan},
  {Zdziarski}, {Zech}, {Zhu}, {Zorn}, {Zouari}, {{\.Z}ywucka}, {MAGIC
  Collaboration}, {Acciari}, {Ansoldi}, {Antonelli}, {Engels}, {Artero},
  {Asano}, {Baack}, {Babi{\'c}}, {Baquero}, {de Almeida}, {Barrio}, {Becerra
  Gonz{\'a}lez}, {Bednarek}, {Bellizzi}, {Bernardini}, {Bernardos}, {Berti},
  {Besenrieder}, {Bhattacharyya}, {Bigongiari}, {Biland}, {Blanch}, {Bonnoli},
  {Bo{\v{s}}njak}, {Busetto}, {Carosi}, {Ceribella}, {Cerruti}, {Chai},
  {Chilingarian}, {Cikota}, {Colak}, {Colombo}, {Contreras}, {Cortina},
  {Covino}, {D'Amico}, {D'Elia}, {da Vela}, {Dazzi}, {de Angelis}, {de Lotto},
  {Delfino}, {Delgado}, {Delgado Mendez}, {Depaoli}, {di Pierro}, {di Venere},
  {Do Souto Espi{\~n}eira}, {Dominis Prester}, {Donini}, {Dorner}, {Doro},
  {Elsaesser}, {Ramazani}, {Fattorini}, {Ferrara}, {Fonseca}, {Font}, {Fruck},
  {Fukami}, {Garc{\'\i}a L{\'o}pez}, {Garczarczyk}, {Gasparyan}, {Gaug},
  {Giglietto}, {Giordano}, {Gliwny}, {Godinovi{\'c}}, {Green}, {Green},
  {Hadasch}, {Hahn}, {Heckmann}, {Herrera}, {Hoang}, {Hrupec}, {H{\"u}tten},
  {Inada}, {Inoue}, {Ishio}, {Iwamura}, {Jim{\'e}nez}, {Jormanainen}, {Jouvin},
  {Kajiwara}, {Karjalainen}, {Kerszberg}, {Kobayashi}, {Kubo}, {Kushida},
  {Lamastra}, {Lelas}, {Leone}, {Lindfors}, {Lombardi}, {Longo},
  {L{\'o}pez-Coto}, {L{\'o}pez-Moya}, {L{\'o}pez-Oramas}, {Loporchio}, {Machado
  de Oliveira Fraga}, {Maggio}, {Majumdar}, {Makariev}, {Mallamaci}, {Maneva},
  {Manganaro}, {Mannheim}, {Maraschi}, {Mariotti}, {Mart{\'\i}nez}, {Mazin},
  {Menchiari}, {Mender}, {Mi{\'c}anovi{\'c}}, {Miceli}, {Miener}, {Minev},
  {Miranda}, {Mirzoyan}, {Molina}, {Moralejo}, {Morcuende}, {Moreno},
  {Moretti}, {Neustroev}, {Nigro}, {Nilsson}, {Nishijima}, {Noda}, {Nozaki},
  {Ohtani}, {Oka}, {Otero-Santos}, {Paiano}, {Palatiello}, {Paneque},
  {Paoletti}, {Paredes}, {Pavleti{\'c}}, {Pe{\~n}il}, {Perennes}, {Persic},
  {Moroni}, {Prandini}, {Priyadarshi}, {Puljak}, {Rhode}, {Rib{\'o}}, {Rico},
  {Righi}, {Rugliancich}, {Saha}, {Sahakyan}, {Saito}, {Sakurai}, {Satalecka},
  {Saturni}, {Schleicher}, {Schmidt}, {Schweizer}, {Sitarek},
  {{\v{S}}nidari{\'c}}, {Sobczynska}, {Spolon}, {Stamerra}, {Strom}, {Strzys},
  {Suda}, {Suri{\'c}}, {Takahashi}, {Tavecchio}, {Temnikov}, {Terzi{\'c}},
  {Teshima}, {Tosti}, {Truzzi}, {Tutone}, {Ubach}, {van Scherpenberg}, {Vanzo},
  {Vazquez Acosta}, {Ventura}, {Verguilov}, {Vigorito}, {Vitale}, {Vovk},
  {Will}, {Wunderlich}, {Zari{\'c}}, {VERITAS Collaboration}, {Adams},
  {Benbow}, {Brill}, {Capasso}, {Christiansen}, {Chromey}, {Daniel}, {Errando},
  {Farrell}, {Feng}, {Finley}, {Fortson}, {Furniss}, {Gent}, {Giuri}, {Hassan},
  {Hervet}, {Holder}, {Hughes}, {Humensky}, {Jin}, {Kaaret}, {Kertzman},
  {Kieda}, {Kumar}, {Lang}, {Lundy}, {Maier}, {Moriarty}, {Mukherjee}, {Nieto},
  {Nievas-Rosillo}, {O'Brien}, {Ong}, {Otte}, {Patel}, {Pfrang}, {Pohl},
  {Prado}, {Pueschel}, {Quinn}, {Ragan}, {Reynolds}, {Ribeiro}, {Richards},
  {Roache}, {Rulten}, {Ryan}, {Santander}, {Sembroski}, {Shang}, {Weinstein},
  {Williams}, {Williamson}, {Eavn Collaboration}, {Hirota}, {Cui}, {Niinuma},
  {Ro}, {Sakai}, {Sawada-Satoh}, {Wajima}, {Wang}, {Liu}, and
  {Yonekura}]{EHT2021}
{EHT MWL Science Working Group}.; {Algaba}, J.C.; {Anczarski}, J.; {Asada}, K.;
  {Balokovi{\'c}}, M.; {Chandra}, S.; {Cui}, Y.Z.; {Falcone}, A.D.;
  {Giroletti}, M.; {Goddi}, C.;  et~al.
\newblock {Broadband Multi-wavelength Properties of M87 during the 2017 Event
  Horizon Telescope Campaign}.
\newblock {\em \apjl} {\bf 2021}, {\em 911},~L11,
  \href{http://xxx.lanl.gov/abs/2104.06855}{{\normalfont
  [arXiv:astro-ph.HE/2104.06855]}}.
\newblock {\url{https://doi.org/10.3847/2041-8213/abef71}}.

\bibitem[{Marscher} \em{et~al.}(2002){Marscher}, {Jorstad}, {G{\'o}mez},
  {Aller}, {Ter{\"a}sranta}, {Lister}, and {Stirling}]{Marscher2002}
{Marscher}, A.P.; {Jorstad}, S.G.; {G{\'o}mez}, J.L.; {Aller}, M.F.;
  {Ter{\"a}sranta}, H.; {Lister}, M.L.; {Stirling}, A.M.
\newblock {Observational evidence for the accretion-disk origin for a radio jet
  in an active galaxy}.
\newblock {\em \nat} {\bf 2002}, {\em 417},~625--627.
\newblock {\url{https://doi.org/10.1038/nature00772}}.

\bibitem[{Chatterjee} \em{et~al.}(2009){Chatterjee}, {Marscher}, {Jorstad},
  {Olmstead}, {McHardy}, {Aller}, {Aller}, {L{\"a}hteenm{\"a}ki}, {Tornikoski},
  {Hovatta}, {Marshall}, {Miller}, {Ryle}, {Chicka}, {Benker}, {Bottorff},
  {Brokofsky}, {Campbell}, {Chonis}, {Gaskell}, {Gaynullina}, {Grankin},
  {Hedrick}, {Ibrahimov}, {Klimek}, {Kruse}, {Masatoshi}, {Miller}, {Pan},
  {Petersen}, {Peterson}, {Shen}, {Strel'nikov}, {Tao}, {Watkins}, and
  {Wheeler}]{Ritaban2009}
{Chatterjee}, R.; {Marscher}, A.P.; {Jorstad}, S.G.; {Olmstead}, A.R.;
  {McHardy}, I.M.; {Aller}, M.F.; {Aller}, H.D.; {L{\"a}hteenm{\"a}ki}, A.;
  {Tornikoski}, M.; {Hovatta}, T.;  et~al.
\newblock {Disk-Jet Connection in the Radio Galaxy 3C 120}.
\newblock {\em \apj} {\bf 2009}, {\em 704},~1689--1703,
  \href{http://xxx.lanl.gov/abs/0909.2051}{{\normalfont
  [arXiv:astro-ph.HE/0909.2051]}}.
\newblock {\url{https://doi.org/10.1088/0004-637X/704/2/1689}}.

\bibitem[{Lohfink} \em{et~al.}(2013){Lohfink}, {Reynolds}, {Jorstad},
  {Marscher}, {Miller}, {Aller}, {Aller}, {Brenneman}, {Fabian}, {Miller},
  {Mushotzky}, {Nowak}, and {Tombesi}]{Lohfink2013}
{Lohfink}, A.M.; {Reynolds}, C.S.; {Jorstad}, S.G.; {Marscher}, A.P.; {Miller},
  E.D.; {Aller}, H.; {Aller}, M.F.; {Brenneman}, L.W.; {Fabian}, A.C.;
  {Miller}, J.M.;  et~al.
\newblock {An X-Ray View of the Jet Cycle in the Radio-loud AGN 3C120}.
\newblock {\em \apj} {\bf 2013}, {\em 772},~83,
  \href{http://xxx.lanl.gov/abs/1305.4937}{{\normalfont
  [arXiv:astro-ph.HE/1305.4937]}}.
\newblock {\url{https://doi.org/10.1088/0004-637X/772/2/83}}.

\bibitem[{Chatterjee} \em{et~al.}(2011){Chatterjee}, {Marscher}, {Jorstad},
  {Markowitz}, {Rivers}, {Rothschild}, {McHardy}, {Aller}, {Aller},
  {L{\"a}hteenm{\"a}ki}, {Tornikoski}, {Harrison}, {Agudo}, {G{\'o}mez},
  {Taylor}, and {Gurwell}]{Ritaban2011}
{Chatterjee}, R.; {Marscher}, A.P.; {Jorstad}, S.G.; {Markowitz}, A.; {Rivers},
  E.; {Rothschild}, R.E.; {McHardy}, I.M.; {Aller}, M.F.; {Aller}, H.D.;
  {L{\"a}hteenm{\"a}ki}, A.;  et~al.
\newblock {Connection Between the Accretion Disk and Jet in the Radio Galaxy 3C
  111}.
\newblock {\em \apj} {\bf 2011}, {\em 734},~43,
  \href{http://xxx.lanl.gov/abs/1104.0663}{{\normalfont
  [arXiv:astro-ph.HE/1104.0663]}}.
\newblock {\url{https://doi.org/10.1088/0004-637X/734/1/43}}.

\bibitem[{Tombesi} \em{et~al.}(2012){Tombesi}, {Sambruna}, {Marscher},
  {Jorstad}, {Reynolds}, and {Markowitz}]{Tombesi2012}
{Tombesi}, F.; {Sambruna}, R.M.; {Marscher}, A.P.; {Jorstad}, S.G.; {Reynolds},
  C.S.; {Markowitz}, A.
\newblock {Comparison of ejection events in the jet and accretion disc outflows
  in 3C 111}.
\newblock {\em \mnras} {\bf 2012}, {\em 424},~754--761,
  \href{http://xxx.lanl.gov/abs/1205.1734}{{\normalfont
  [arXiv:astro-ph.HE/1205.1734]}}.
\newblock {\url{https://doi.org/10.1111/j.1365-2966.2012.21266.x}}.

\bibitem[{Weaver} \em{et~al.}(2022){Weaver}, {Jorstad}, {Marscher}, {Morozova},
  {Troitsky}, {Agudo}, {G{\'o}mez}, {L{\"a}hteenm{\"a}ki}, {Tammi}, and
  {Tornikoski}]{Weaver2022}
{Weaver}, Z.R.; {Jorstad}, S.G.; {Marscher}, A.P.; {Morozova}, D.A.;
  {Troitsky}, I.S.; {Agudo}, I.; {G{\'o}mez}, J.L.; {L{\"a}hteenm{\"a}ki}, A.;
  {Tammi}, J.; {Tornikoski}, M.
\newblock {Kinematics of Parsec-scale Jets of Gamma-Ray Blazars at 43 GHz
  during 10 yr of the VLBA-BU-BLAZAR Program}.
\newblock {\em \apjs} {\bf 2022}, {\em 260},~12,
  \href{http://xxx.lanl.gov/abs/2202.12290}{{\normalfont
  [arXiv:astro-ph.HE/2202.12290]}}.
\newblock {\url{https://doi.org/10.3847/1538-4365/ac589c}}.

\bibitem[{Fender} \em{et~al.}(2004){Fender}, {Belloni}, and
  {Gallo}]{Fender2004}
{Fender}, R.P.; {Belloni}, T.M.; {Gallo}, E.
\newblock {Towards a unified model for black hole X-ray binary jets}.
\newblock {\em \mnras} {\bf 2004}, {\em 355},~1105--1118,
  \href{http://xxx.lanl.gov/abs/astro-ph/0409360}{{\normalfont
  [arXiv:astro-ph/astro-ph/0409360]}}.
\newblock {\url{https://doi.org/10.1111/j.1365-2966.2004.08384.x}}.

\bibitem[{Kovalev} \em{et~al.}(2020){Kovalev}, {Pushkarev}, {Nokhrina},
  {Plavin}, {Beskin}, {Chernoglazov}, {Lister}, and {Savolainen}]{Kovalev2020}
{Kovalev}, Y.Y.; {Pushkarev}, A.B.; {Nokhrina}, E.E.; {Plavin}, A.V.; {Beskin},
  V.S.; {Chernoglazov}, A.V.; {Lister}, M.L.; {Savolainen}, T.
\newblock {A transition from parabolic to conical shape as a common effect in
  nearby AGN jets}.
\newblock {\em \mnras} {\bf 2020}, {\em 495},~3576--3591,
  \href{http://xxx.lanl.gov/abs/1907.01485}{{\normalfont
  [arXiv:astro-ph.GA/1907.01485]}}.
\newblock {\url{https://doi.org/10.1093/mnras/staa1121}}.

\bibitem[{Casadio} \em{et~al.}(2021){Casadio}, {MacDonald}, {Boccardi},
  {Jorstad}, {Marscher}, {Krichbaum}, {Hodgson}, {Kim}, {Traianou}, {Weaver},
  {G{\'o}mez Garrido}, {Gonz{\'a}lez Garc{\'\i}a}, {Kallunki}, {Lindqvist},
  {S{\'a}nchez}, {Yang}, and {Zensus}]{Casadio2021}
{Casadio}, C.; {MacDonald}, N.R.; {Boccardi}, B.; {Jorstad}, S.G.; {Marscher},
  A.P.; {Krichbaum}, T.P.; {Hodgson}, J.A.; {Kim}, J.Y.; {Traianou}, E.;
  {Weaver}, Z.R.;  et~al.
\newblock {The jet collimation profile at high resolution in BL Lacertae}.
\newblock {\em \aap} {\bf 2021}, {\em 649},~A153,
  \href{http://xxx.lanl.gov/abs/2102.08952}{{\normalfont
  [arXiv:astro-ph.HE/2102.08952]}}.
\newblock {\url{https://doi.org/10.1051/0004-6361/202039616}}.

\bibitem[{Marscher} \em{et~al.}(2008){Marscher}, {Jorstad}, {D'Arcangelo},
  {Smith}, {Williams}, {Larionov}, {Oh}, {Olmstead}, {Aller}, {Aller},
  {McHardy}, {L{\"a}hteenm{\"a}ki}, {Tornikoski}, {Valtaoja}, {Hagen-Thorn},
  {Kopatskaya}, {Gear}, {Tosti}, {Kurtanidze}, {Nikolashvili}, {Sigua},
  {Miller}, and {Ryle}]{Marscher2008}
{Marscher}, A.P.; {Jorstad}, S.G.; {D'Arcangelo}, F.D.; {Smith}, P.S.;
  {Williams}, G.G.; {Larionov}, V.M.; {Oh}, H.; {Olmstead}, A.R.; {Aller},
  M.F.; {Aller}, H.D.;  et~al.
\newblock {The inner jet of an active galactic nucleus as revealed by a
  radio-to-{\ensuremath{\gamma}}-ray outburst}.
\newblock {\em \nat} {\bf 2008}, {\em 452},~966--969.
\newblock {\url{https://doi.org/10.1038/nature06895}}.

\bibitem[{Liodakis} \em{et~al.}(2022){Liodakis}, {Marscher}, {Agudo},
  {Berdyugin}, {Bernardos}, {Bonnoli}, {Borman}, {Casadio}, {Casanova},
  {Cavazzuti}, {Cavero}, {Di Gesu}, {Di Lalla}, {Donnarumma}, {Ehlert},
  {Errando}, {Escudero}, {Garc{\'\i}a-Comas}, {Ag{\'\i}s-Gonz{\'a}lez},
  {Husillos}, {Jormanainen}, {Jorstad}, {Kagitani}, {Kopatskaya}, {Kravtsov},
  {Krawczynski}, {Lindfors}, {Larionova}, {Madejski}, {Marin}, {Marchini},
  {Marshall}, {Morozova}, {Massaro}, {Masiero}, {Mawet}, {Middei},
  {Millar-Blanchaer}, {Myserlis}, {Negro}, {Nilsson}, {O'Dell}, {Omodei},
  {Pacciani}, {Paggi}, {Panopoulou}, {Peirson}, {Perri}, {Petrucci},
  {Poutanen}, {Puccetti}, {Romani}, {Sakanoi}, {Savchenko}, {Sota},
  {Tavecchio}, {Tinyanont}, {Vasiliev}, {Weaver}, {Zhovtan}, {Antonelli},
  {Bachetti}, {Baldini}, {Baumgartner}, {Bellazzini}, {Bianchi}, {Bongiorno},
  {Bonino}, {Brez}, {Bucciantini}, {Capitanio}, {Castellano}, {Ciprini},
  {Costa}, {De Rosa}, {Del Monte}, {Di Marco}, {Doroshenko}, {Dov{\v{c}}iak},
  {Enoto}, {Evangelista}, {Fabiani}, {Ferrazzoli}, {Garcia}, {Gunji},
  {Hayashida}, {Heyl}, {Iwakiri}, {Karas}, {Kitaguchi}, {Kolodziejczak}, {La
  Monaca}, {Latronico}, {Maldera}, {Manfreda}, {Marinucci}, {Matt},
  {Mitsuishi}, {Mizuno}, {Muleri}, {Ng}, {Oppedisano}, {Papitto}, {Pavlov},
  {Pesce-Rollins}, {Pilia}, {Possenti}, {Ramsey}, {Rankin}, {Ratheesh},
  {Sgr{\'o}}, {Slane}, {Soffitta}, {Spandre}, {Tamagawa}, {Taverna}, {Tawara},
  {Tennant}, {Thomas}, {Tombesi}, {Trois}, {Tsygankov}, {Turolla}, {Vink},
  {Weisskopf}, {Wu}, {Xie}, and {Zane}]{Liodakis2022}
{Liodakis}, I.; {Marscher}, A.P.; {Agudo}, I.; {Berdyugin}, A.V.; {Bernardos},
  M.I.; {Bonnoli}, G.; {Borman}, G.A.; {Casadio}, C.; {Casanova}, V.;
  {Cavazzuti}, E.;  et~al.
\newblock {Polarized Blazar X-rays imply particle acceleration in shocks}.
\newblock {\em arXiv e-prints} {\bf 2022}, p. arXiv:2209.06227,
  \href{http://xxx.lanl.gov/abs/2209.06227}{{\normalfont
  [arXiv:astro-ph.HE/2209.06227]}}.

\bibitem[{Kim} \em{et~al.}(2020){Kim}, {Krichbaum}, {Broderick}, {Wielgus},
  {Blackburn}, {G{\'o}mez}, {Johnson}, {Bouman}, {Chael}, {Akiyama}, {Jorstad},
  {Marscher}, {Issaoun}, {Janssen}, {Chan}, {Savolainen}, {Pesce}, {{\"O}zel},
  {Alberdi}, {Alef}, {Asada}, {Azulay}, {Baczko}, {Ball}, {Balokovi{\'c}},
  {Barrett}, {Bintley}, {Boland}, {Bower}, {Bremer}, {Brinkerink},
  {Brissenden}, {Britzen}, {Broguiere}, {Bronzwaer}, {Byun}, {Carlstrom},
  {Chatterjee}, {Chatterjee}, {Chen}, {Chen}, {Cho}, {Christian}, {Conway},
  {Cordes}, {Crew}, {Cui}, {Davelaar}, {De Laurentis}, {Deane}, {Dempsey},
  {Desvignes}, {Dexter}, {Doeleman}, {Eatough}, {Falcke}, {Fish}, {Fomalont},
  {Fraga-Encinas}, {Friberg}, {Fromm}, {Galison}, {Gammie}, {Garc{\'\i}a},
  {Gentaz}, {Georgiev}, {Goddi}, {Gold}, {G{\'o}mez-Ruiz}, {Gu}, {Gurwell},
  {Hada}, {Hecht}, {Hesper}, {Ho}, {Ho}, {Honma}, {Huang}, {Huang}, {Hughes},
  {Ikeda}, {Inoue}, {James}, {Jannuzi}, {Jeter}, {Jiang}, {Jimenez-Rosales},
  {Jung}, {Karami}, {Karuppusamy}, {Kawashima}, {Keating}, {Kettenis}, {Kim},
  {Kim}, {Kino}, {Koay}, {Koch}, {Koyama}, {Kramer}, {Kramer}, {Kuo}, {Lauer},
  {Lee}, {Li}, {Li}, {Lindqvist}, {Lico}, {Liu}, {Liuzzo}, {Lo}, {Lobanov},
  {Loinard}, {Lonsdale}, {Lu}, {MacDonald}, {Mao}, {Markoff}, {Marrone},
  {Mart{\'\i}-Vidal}, {Matsushita}, {Matthews}, {Medeiros}, {Menten}, {Mizuno},
  {Mizuno}, {Moran}, {Moriyama}, {Moscibrodzka}, {Musoke}, {M{\"u}ller},
  {Nagai}, {Nagar}, {Nakamura}, {Narayan}, {Narayanan}, {Natarajan}, {Neri},
  {Ni}, {Noutsos}, {Okino}, {Olivares}, {Ortiz-Le{\'o}n}, {Oyama}, {Palumbo},
  {Park}, {Patel}, {Pen}, {Pi{\'e}tu}, {Plambeck}, {PopStefanija}, {Porth},
  {Prather}, {Preciado-L{\'o}pez}, {Psaltis}, {Pu}, {Ramakrishnan}, {Rao},
  {Rawlings}, {Raymond}, {Rezzolla}, {Ripperda}, {Roelofs}, {Rogers}, {Ros},
  {Rose}, {Roshanineshat}, {Rottmann}, {Roy}, {Ruszczyk}, {Ryan}, {Rygl},
  {S{\'a}nchez}, {S{\'a}nchez-Arguelles}, {Sasada}, {Schloerb}, {Schuster},
  {Shao}, {Shen}, {Small}, {Sohn}, {SooHoo}, {Tazaki}, {Tiede}, {Tilanus},
  {Titus}, {Toma}, {Torne}, {Trent}, {Traianou}, {Trippe}, {Tsuda}, {van
  Bemmel}, {van Langevelde}, {van Rossum}, {Wagner}, {Wardle}, {Ward-Thompson},
  {Weintroub}, {Wex}, {Wharton}, {Wong}, {Wu}, {Yoon}, {Young}, {Young},
  {Younsi}, {Yuan}, {Yuan}, {Zensus}, {Zhao}, {Zhao}, {Zhu}, {Algaba},
  {Allardi}, {Amestica}, {Anczarski}, {Bach}, {Baganoff}, {Beaudoin}, {Benson},
  {Berthold}, {Blanchard}, {Blundell}, {Bustamente}, {Cappallo},
  {Castillo-Dom{\'\i}nguez}, {Chang}, {Chang}, {Chang}, {Chen}, {Chilson},
  {Chuter}, {Rosado}, {Coulson}, {Crowley}, {Derome}, {Dexter}, {Dornbusch},
  {Dudevoir}, {Dzib}, {Eckart}, {Eckert}, {Erickson}, {Everett}, {Faber},
  {Farah}, {Fath}, {Folkers}, {Forbes}, {Freund}, {Gale}, {Gao}, {Geertsema},
  {Graham}, {Greer}, {Grosslein}, {Gueth}, {Haggard}, {Halverson}, {Han},
  {Han}, {Hao}, {Hasegawa}, {Henning}, {Hern{\'a}ndez-G{\'o}mez},
  {Herrero-Illana}, {Heyminck}, {Hirota}, {Hoge}, {Huang}, {Violette
  Impellizzeri}, {Jiang}, {John}, {Kamble}, {Keisler}, {Kimura}, {Kono},
  {Kubo}, {Kuroda}, {Lacasse}, {Laing}, {Leitch}, {Li}, {Lin}, {Liu}, {Liu},
  {Lu}, {Marson}, {Martin-Cocher}, {Massingill}, {Matulonis}, {McColl},
  {McWhirter}, {Messias}, {Meyer-Zhao}, {Michalik}, {Monta{\~n}a},
  {Montgomerie}, {Mora-Klein}, {Muders}, {Nadolski}, {Navarro}, {Neilsen},
  {Nguyen}, {Nishioka}, {Norton}, {Nowak}, {Nystrom}, {Ogawa}, {Oshiro},
  {Oyama}, {Parsons}, {Pe{\~n}alver}, {Phillips}, {Poirier}, {Pradel},
  {Primiani}, {Raffin}, {Rahlin}, {Reiland}, {Risacher}, {Ruiz},
  {S{\'a}ez-Mada{\'\i}n}, {Sassella}, {Schellart}, {Shaw}, {Silva}, {Shiokawa},
  {Smith}, {Snow}, {Souccar}, {Sousa}, {Sridharan}, {Srinivasan}, {Stahm},
  {Stark}, {Story}, {Timmer}, {Vertatschitsch}, {Walther}, {Wei}, {Whitehorn},
  {Whitney}, {Woody}, {Wouterloot}, {Wright}, {Yamaguchi}, {Yu}, {Zeballos},
  {Zhang}, {Ziurys}, and {Event Horizon Telescope Collaboration}]{kim2020}
{Kim}, J.Y.; {Krichbaum}, T.P.; {Broderick}, A.E.; {Wielgus}, M.; {Blackburn},
  L.; {G{\'o}mez}, J.L.; {Johnson}, M.D.; {Bouman}, K.L.; {Chael}, A.;
  {Akiyama}, K.;  et~al.
\newblock {Event Horizon Telescope imaging of the archetypal blazar 3C 279 at
  an extreme 20 microarcsecond resolution}.
\newblock {\em \aap} {\bf 2020}, {\em 640},~A69.
\newblock {\url{https://doi.org/10.1051/0004-6361/202037493}}.

\bibitem[{Cawthorne} \em{et~al.}(2013){Cawthorne}, {Jorstad}, and
  {Marscher}]{Cawthorne2013}
{Cawthorne}, T.V.; {Jorstad}, S.G.; {Marscher}, A.P.
\newblock {Polarization Structure in the Core of 1803+784: A Signature of
  Recollimation Shocks?}
\newblock {\em \apj} {\bf 2013}, {\em 772},~14,
  \href{http://xxx.lanl.gov/abs/1305.5356}{{\normalfont
  [arXiv:astro-ph.HE/1305.5356]}}.
\newblock {\url{https://doi.org/10.1088/0004-637X/772/1/14}}.

\bibitem[{Gabuzda} \em{et~al.}(2015){Gabuzda}, {Knuettel}, and
  {Reardon}]{Gabuzda2015}
{Gabuzda}, D.C.; {Knuettel}, S.; {Reardon}, B.
\newblock {Transverse Faraday-rotation gradients across the jets of 15 active
  galactic nuclei}.
\newblock {\em \mnras} {\bf 2015}, {\em 450},~2441--2450,
  \href{http://xxx.lanl.gov/abs/1503.03411}{{\normalfont
  [arXiv:astro-ph.GA/1503.03411]}}.
\newblock {\url{https://doi.org/10.1093/mnras/stv555}}.

\bibitem[{Lico} \em{et~al.}(2012){Lico}, {Giroletti}, {Orienti}, {Giovannini},
  {Cotton}, {Edwards}, {Fuhrmann}, {Krichbaum}, {Sokolovsky}, {Kovalev},
  {Jorstad}, {Marscher}, {Kino}, {Paneque}, {Perez-Torres}, and
  {Piner}]{Lico2012}
{Lico}, R.; {Giroletti}, M.; {Orienti}, M.; {Giovannini}, G.; {Cotton}, W.;
  {Edwards}, P.G.; {Fuhrmann}, L.; {Krichbaum}, T.P.; {Sokolovsky}, K.V.;
  {Kovalev}, Y.Y.;  et~al.
\newblock {VLBA monitoring of Mrk 421 at 15 GHz and 24 GHz during 2011}.
\newblock {\em \aap} {\bf 2012}, {\em 545},~A117,
  \href{http://xxx.lanl.gov/abs/1208.5853}{{\normalfont
  [arXiv:astro-ph.HE/1208.5853]}}.
\newblock {\url{https://doi.org/10.1051/0004-6361/201219870}}.

\bibitem[{Ghisellini} \em{et~al.}(2005){Ghisellini}, {Tavecchio}, and
  {Chiaberge}]{Ghisellini2005}
{Ghisellini}, G.; {Tavecchio}, F.; {Chiaberge}, M.
\newblock {Structured jets in TeV BL Lac objects and radiogalaxies.
  Implications for the observed properties}.
\newblock {\em \aap} {\bf 2005}, {\em 432},~401--410,
  \href{http://xxx.lanl.gov/abs/astro-ph/0406093}{{\normalfont
  [arXiv:astro-ph/astro-ph/0406093]}}.
\newblock {\url{https://doi.org/10.1051/0004-6361:20041404}}.

\bibitem[{Giroletti} \em{et~al.}(2008){Giroletti}, {Giovannini}, {Cotton},
  {Taylor}, {P{\'e}rez-Torres}, {Chiaberge}, and {Edwards}]{Giroletti2008}
{Giroletti}, M.; {Giovannini}, G.; {Cotton}, W.D.; {Taylor}, G.B.;
  {P{\'e}rez-Torres}, M.A.; {Chiaberge}, M.; {Edwards}, P.G.
\newblock {The jet of Markarian 501 from millions of Schwarzschild radii down
  to a few hundreds}.
\newblock {\em \aap} {\bf 2008}, {\em 488},~905--914,
  \href{http://xxx.lanl.gov/abs/0807.1786}{{\normalfont
  [arXiv:astro-ph/0807.1786]}}.
\newblock {\url{https://doi.org/10.1051/0004-6361:200809784}}.

\bibitem[{Lico} \em{et~al.}(2014){Lico}, {Giroletti}, {Orienti}, {G{\'o}mez},
  {Casadio}, {D'Ammando}, {Blasi}, {Cotton}, {Edwards}, {Fuhrmann}, {Jorstad},
  {Kino}, {Kovalev}, {Krichbaum}, {Marscher}, {Paneque}, {Piner}, and
  {Sokolovsky}]{Lico2014}
{Lico}, R.; {Giroletti}, M.; {Orienti}, M.; {G{\'o}mez}, J.L.; {Casadio}, C.;
  {D'Ammando}, F.; {Blasi}, M.G.; {Cotton}, W.; {Edwards}, P.G.; {Fuhrmann},
  L.;  et~al.
\newblock {Very Long Baseline polarimetry and the {\ensuremath{\gamma}}-ray
  connection in Markarian 421 during the broadband campaign in 2011}.
\newblock {\em \aap} {\bf 2014}, {\em 571},~A54,
  \href{http://xxx.lanl.gov/abs/1410.0884}{{\normalfont
  [arXiv:astro-ph.HE/1410.0884]}}.
\newblock {\url{https://doi.org/10.1051/0004-6361/201424341}}.

\bibitem[{Koyama} \em{et~al.}(2019){Koyama}, {Kino}, {Doi}, {Niinuma},
  {Giroletti}, {Paneque}, {Akiyama}, {Giovannini}, {Zhao}, {Ros}, {Kataoka},
  {Orienti}, {Hada}, {Nagai}, {Isobe}, {Kobayashi}, {Honma}, and
  {Lico}]{Koyama2019}
{Koyama}, S.; {Kino}, M.; {Doi}, A.; {Niinuma}, K.; {Giroletti}, M.; {Paneque},
  D.; {Akiyama}, K.; {Giovannini}, G.; {Zhao}, G.Y.; {Ros}, E.;  et~al.
\newblock {Stable Radio Core of the Blazar Mrk 501 during High-energy Active
  State in 2012}.
\newblock {\em \apj} {\bf 2019}, {\em 884},~132.
\newblock {\url{https://doi.org/10.3847/1538-4357/ab4260}}.

\bibitem[{Hovatta} \em{et~al.}(2012){Hovatta}, {Lister}, {Aller}, {Aller},
  {Homan}, {Kovalev}, {Pushkarev}, and {Savolainen}]{Hovatta2012}
{Hovatta}, T.; {Lister}, M.L.; {Aller}, M.F.; {Aller}, H.D.; {Homan}, D.C.;
  {Kovalev}, Y.Y.; {Pushkarev}, A.B.; {Savolainen}, T.
\newblock {MOJAVE: Monitoring of Jets in Active Galactic Nuclei with VLBA
  Experiments. VIII. Faraday Rotation in Parsec-scale AGN Jets}.
\newblock {\em \aj} {\bf 2012}, {\em 144},~105,
  \href{http://xxx.lanl.gov/abs/1205.6746}{{\normalfont
  [arXiv:astro-ph.CO/1205.6746]}}.
\newblock {\url{https://doi.org/10.1088/0004-6256/144/4/105}}.

\bibitem[{Nagai} \em{et~al.}(2014){Nagai}, {Haga}, {Giovannini}, {Doi},
  {Orienti}, {D'Ammando}, {Kino}, {Nakamura}, {Asada}, {Hada}, and
  {Giroletti}]{Nagai2014}
{Nagai}, H.; {Haga}, T.; {Giovannini}, G.; {Doi}, A.; {Orienti}, M.;
  {D'Ammando}, F.; {Kino}, M.; {Nakamura}, M.; {Asada}, K.; {Hada}, K.;  et~al.
\newblock {Limb-brightened Jet of 3C 84 Revealed by the 43 GHz
  Very-Long-Baseline-Array Observation}.
\newblock {\em \apj} {\bf 2014}, {\em 785},~53,
  \href{http://xxx.lanl.gov/abs/1402.5930}{{\normalfont
  [arXiv:astro-ph.HE/1402.5930]}}.
\newblock {\url{https://doi.org/10.1088/0004-637X/785/1/53}}.

\bibitem[{Janssen} \em{et~al.}(2021){Janssen}, {Falcke}, {Kadler}, {Ros},
  {Wielgus}, {Akiyama}, {Balokovi{\'c}}, {Blackburn}, {Bouman}, {Chael},
  {Chan}, {Chatterjee}, {Davelaar}, {Edwards}, {Fromm}, {G{\'o}mez}, {Goddi},
  {Issaoun}, {Johnson}, {Kim}, {Koay}, {Krichbaum}, {Liu}, {Liuzzo}, {Markoff},
  {Markowitz}, {Marrone}, {Mizuno}, {M{\"u}ller}, {Ni}, {Pesce},
  {Ramakrishnan}, {Roelofs}, {Rygl}, {van Bemmel}, {Event Horizon Telescope
  Collaboration}, {Alberdi}, {Alef}, {Algaba}, {Anantua}, {Asada}, {Azulay},
  {Baczko}, {Ball}, {Ball}, {Barrett}, {Benson}, {Bintley}, {Bintley},
  {Blundell}, {Boland}, {Boland}, {Bower}, {Boyce}, {Bremer}, {Brinkerink},
  {Brissenden}, {Britzen}, {Broderick}, {Broguiere}, {Bronzwaer}, {Byun},
  {Carlstrom}, {Chatterjee}, {Chen}, {Chen}, {Chesler}, {Cho}, {Christian},
  {Conway}, {Cordes}, {Crawford}, {Crew}, {Cruz-Osorio}, {Cui}, {Cui}, {De
  Laurentis}, {Deane}, {Dempsey}, {Desvignes}, {Dexter}, {Doeleman}, {Eatough},
  {Farah}, {Farah}, {Fish}, {Fomalont}, {Ford}, {Fraga-Encinas}, {Friberg},
  {Friberg}, {Fuentes}, {Galison}, {Gammie}, {Garc{\'\i}a}, {Gelles}, {Gentaz},
  {Georgiev}, {Georgiev}, {Gold}, {Gold}, {G{\'o}mez-Ruiz}, {Gu}, {Gurwell},
  {Hada}, {Haggard}, {Hecht}, {Hesper}, {Himwich}, {Ho}, {Ho}, {Honma},
  {Huang}, {Huang}, {Hughes}, {Ikeda}, {Inoue}, {Inoue}, {James}, {Jannuzi},
  {Jannuzi}, {Jeter}, {Jiang}, {Jimenez-Rosales}, {Jimenez-Rosales}, {Jorstad},
  {Jung}, {Karami}, {Karuppusamy}, {Kawashima}, {Keating}, {Kettenis}, {Kim},
  {Kim}, {Kim}, {Kim}, {Kino}, {Kino}, {Kofuji}, {Koyama}, {Kramer}, {Kramer},
  {Kramer}, {Kuo}, {Lauer}, {Lee}, {Levis}, {Li}, {Li}, {Lindqvist}, {Lico},
  {Lindahl}, {Lindahl}, {Liu}, {Liu}, {Lo}, {Lobanov}, {Loinard}, {Lonsdale},
  {Lu}, {MacDonald}, {Mao}, {Marchili}, {Marchili}, {Marchili}, {Marscher},
  {Mart{\'\i}-Vidal}, {Matsushita}, {Matthews}, {Medeiros}, {Menten}, {Mizuno},
  {Mizuno}, {Moran}, {Moriyama}, {Moscibrodzka}, {Moscibrodzka}, {Musoke},
  {Mej{\'\i}as}, {Nagai}, {Nagar}, {Nakamura}, {Narayan}, {Narayanan},
  {Natarajan}, {Nathanail}, {Neilsen}, {Neri}, {Neri}, {Noutsos}, {Nowak},
  {Okino}, {Olivares}, {Ortiz-Le{\'o}n}, {Oyama}, {{\"O}zel}, {Palumbo},
  {Park}, {Patel}, {Pen}, {Pen}, {Pi{\'e}tu}, {Plambeck}, {PopStefanija},
  {Porth}, {P{\"o}tzl}, {Prather}, {Preciado-L{\'o}pez}, {Psaltis}, {Pu}, {Pu},
  {Rao}, {Rawlings}, {Raymond}, {Rezzolla}, {Ricarte}, {Ripperda}, {Ripperda},
  {Rogers}, {Rogers}, {Rose}, {Roshanineshat}, {Rottmann}, {Roy}, {Ruszczyk},
  {Ruszczyk}, {S{\'a}nchez}, {S{\'a}nchez-Arguelles}, {Sasada}, {Savolainen},
  {Schloerb}, {Schuster}, {Shao}, {Shen}, {Small}, {Sohn}, {SooHoo}, {Sun},
  {Tazaki}, {Tetarenko}, {Tiede}, {Tilanus}, {Titus}, {Torne}, {Trent},
  {Traianou}, {Trippe}, {van Bemmel}, {van Langevelde}, {van Rossum}, {Wagner},
  {Ward-Thompson}, {Wardle}, {Weintroub}, {Wex}, {Wharton}, {Wharton}, {Wong},
  {Wu}, {Yoon}, {Young}, {Young}, {Younsi}, {Yuan}, {Yuan}, {Zensus}, {Zhao},
  and {Zhao}]{Janssen2021}
{Janssen}, M.; {Falcke}, H.; {Kadler}, M.; {Ros}, E.; {Wielgus}, M.; {Akiyama},
  K.; {Balokovi{\'c}}, M.; {Blackburn}, L.; {Bouman}, K.L.; {Chael}, A.;
  et~al.
\newblock {Event Horizon Telescope observations of the jet launching and
  collimation in Centaurus A}.
\newblock {\em Nature Astronomy} {\bf 2021}, {\em 5},~1017--1028,
  \href{http://xxx.lanl.gov/abs/2111.03356}{{\normalfont
  [arXiv:astro-ph.GA/2111.03356]}}.
\newblock {\url{https://doi.org/10.1038/s41550-021-01417-w}}.

\bibitem[{Bruni} \em{et~al.}(2021){Bruni}, {G{\'o}mez}, {Vega-Garc{\'\i}a},
  {Lobanov}, {Fuentes}, {Savolainen}, {Kovalev}, {Perucho}, {Mart{\'\i}},
  {Anderson}, {Edwards}, {Gurvits}, {Lisakov}, {Pushkarev}, {Sokolovsky}, and
  {Zensus}]{Bruni2021}
{Bruni}, G.; {G{\'o}mez}, J.L.; {Vega-Garc{\'\i}a}, L.; {Lobanov}, A.P.;
  {Fuentes}, A.; {Savolainen}, T.; {Kovalev}, Y.Y.; {Perucho}, M.;
  {Mart{\'\i}}, J.M.; {Anderson}, J.M.;  et~al.
\newblock {RadioAstron reveals a spine-sheath jet structure in 3C 273}.
\newblock {\em \aap} {\bf 2021}, {\em 654},~A27,
  \href{http://xxx.lanl.gov/abs/2101.07324}{{\normalfont
  [arXiv:astro-ph.GA/2101.07324]}}.
\newblock {\url{https://doi.org/10.1051/0004-6361/202039423}}.

\bibitem[{Valtonen} \em{et~al.}(2008){Valtonen}, {Lehto}, {Nilsson}, {Heidt},
  {Takalo}, {Sillanp{\"a}{\"a}}, {Villforth}, {Kidger}, {Poyner}, {Pursimo},
  {Zola}, {Wu}, {Zhou}, {Sadakane}, {Drozdz}, {Koziel}, {Marchev}, {Ogloza},
  {Porowski}, {Siwak}, {Stachowski}, {Winiarski}, {Hentunen}, {Nissinen},
  {Liakos}, and {Dogru}]{Valtonen2008}
{Valtonen}, M.J.; {Lehto}, H.J.; {Nilsson}, K.; {Heidt}, J.; {Takalo}, L.O.;
  {Sillanp{\"a}{\"a}}, A.; {Villforth}, C.; {Kidger}, M.; {Poyner}, G.;
  {Pursimo}, T.;  et~al.
\newblock {A massive binary black-hole system in OJ287 and a test of general
  relativity}.
\newblock {\em \nat} {\bf 2008}, {\em 452},~851--853,
  \href{http://xxx.lanl.gov/abs/0809.1280}{{\normalfont
  [arXiv:astro-ph/0809.1280]}}.
\newblock {\url{https://doi.org/10.1038/nature06896}}.

\bibitem[{G{\'o}mez} \em{et~al.}(2022){G{\'o}mez}, {Traianou}, {Krichbaum},
  {Lobanov}, {Fuentes}, {Lico}, {Zhao}, {Bruni}, {Kovalev},
  {L{\"a}hteenm{\"a}ki}, {Voitsik}, {Lisakov}, {Angelakis}, {Bach}, {Casadio},
  {Cho}, {Dey}, {Gopakumar}, {Gurvits}, {Jorstad}, {Kovalev}, {Lister},
  {Marscher}, {Myserlis}, {Pushkarev}, {Ros}, {Savolainen}, {Tornikoski},
  {Valtonen}, and {Zensus}]{Gomez2022}
{G{\'o}mez}, J.L.; {Traianou}, E.; {Krichbaum}, T.P.; {Lobanov}, A.P.;
  {Fuentes}, A.; {Lico}, R.; {Zhao}, G.Y.; {Bruni}, G.; {Kovalev}, Y.Y.;
  {L{\"a}hteenm{\"a}ki}, A.;  et~al.
\newblock {Probing the Innermost Regions of AGN Jets and Their Magnetic Fields
  with RadioAstron. V. Space and Ground Millimeter-VLBI Imaging of OJ 287}.
\newblock {\em \apj} {\bf 2022}, {\em 924},~122,
  \href{http://xxx.lanl.gov/abs/2111.11200}{{\normalfont
  [arXiv:astro-ph.HE/2111.11200]}}.
\newblock {\url{https://doi.org/10.3847/1538-4357/ac3bcc}}.

\bibitem[{Liska} \em{et~al.}(2018){Liska}, {Hesp}, {Tchekhovskoy}, {Ingram},
  {van der Klis}, and {Markoff}]{Liska2018}
{Liska}, M.; {Hesp}, C.; {Tchekhovskoy}, A.; {Ingram}, A.; {van der Klis}, M.;
  {Markoff}, S.
\newblock {Formation of precessing jets by tilted black hole discs in 3D
  general relativistic MHD simulations}.
\newblock {\em \mnras} {\bf 2018}, {\em 474},~L81--L85,
  \href{http://xxx.lanl.gov/abs/1707.06619}{{\normalfont
  [arXiv:astro-ph.HE/1707.06619]}}.
\newblock {\url{https://doi.org/10.1093/mnrasl/slx174}}.

\bibitem[{Pesce} \em{et~al.}(2021){Pesce}, {Palumbo}, {Narayan}, {Blackburn},
  {Doeleman}, {Johnson}, {Ma}, {Nagar}, {Natarajan}, and {Ricarte}]{Pesce2021}
{Pesce}, D.W.; {Palumbo}, D.C.M.; {Narayan}, R.; {Blackburn}, L.; {Doeleman},
  S.S.; {Johnson}, M.D.; {Ma}, C.P.; {Nagar}, N.M.; {Natarajan}, P.; {Ricarte},
  A.
\newblock {Toward Determining the Number of Observable Supermassive Black Hole
  Shadows}.
\newblock {\em \apj} {\bf 2021}, {\em 923},~260,
  \href{http://xxx.lanl.gov/abs/2108.05228}{{\normalfont
  [arXiv:astro-ph.HE/2108.05228]}}.
\newblock {\url{https://doi.org/10.3847/1538-4357/ac2eb5}}.

\bibitem[{Porth} \em{et~al.}(2017){Porth}, {Olivares}, {Mizuno}, {Younsi},
  {Rezzolla}, {Moscibrodzka}, {Falcke}, and {Kramer}]{Porth2017}
{Porth}, O.; {Olivares}, H.; {Mizuno}, Y.; {Younsi}, Z.; {Rezzolla}, L.;
  {Moscibrodzka}, M.; {Falcke}, H.; {Kramer}, M.
\newblock {The black hole accretion code}.
\newblock {\em Computational Astrophysics and Cosmology} {\bf 2017}, {\em
  4},~1,  \href{http://xxx.lanl.gov/abs/1611.09720}{{\normalfont
  [arXiv:gr-qc/1611.09720]}}.
\newblock {\url{https://doi.org/10.1186/s40668-017-0020-2}}.

\bibitem[{Giommi} and {Padovani}(2021)]{Giommi2021}
{Giommi}, P.; {Padovani}, P.
\newblock {Astrophysical Neutrinos and Blazars}.
\newblock {\em Universe} {\bf 2021}, {\em 7},~492,
  \href{http://xxx.lanl.gov/abs/2112.06232}{{\normalfont
  [arXiv:astro-ph.HE/2112.06232]}}.
\newblock {\url{https://doi.org/10.3390/universe7120492}}.

\bibitem[{Nanci} \em{et~al.}(2022){Nanci}, {Giroletti}, {Orienti}, {Migliori},
  {Mold{\'o}n}, {Garrappa}, {Kadler}, {Ros}, {Buson}, {An}, {P{\'e}rez-Torres},
  {D'Ammando}, {Mohan}, {Agudo}, {Sohn}, {Castro-Tirado}, and
  {Zhang}]{Nanci2022}
{Nanci}, C.; {Giroletti}, M.; {Orienti}, M.; {Migliori}, G.; {Mold{\'o}n}, J.;
  {Garrappa}, S.; {Kadler}, M.; {Ros}, E.; {Buson}, S.; {An}, T.;  et~al.
\newblock {Observing the inner parsec-scale region of candidate
  neutrino-emitting blazars}.
\newblock {\em \aap} {\bf 2022}, {\em 663},~A129,
  \href{http://xxx.lanl.gov/abs/2203.13268}{{\normalfont
  [arXiv:astro-ph.HE/2203.13268]}}.
\newblock {\url{https://doi.org/10.1051/0004-6361/202142665}}.

\bibitem[{Blinov} \em{et~al.}(2016){Blinov}, {Pavlidou}, {Papadakis},
  {Hovatta}, {Pearson}, {Liodakis}, {Panopoulou}, {Angelakis}, {Balokovi{\'c}},
  {Das}, {Khodade}, {Kiehlmann}, {King}, {Kus}, {Kylafis}, {Mahabal},
  {Marecki}, {Modi}, {Myserlis}, {Paleologou}, {Papamastorakis}, {Pazderska},
  {Pazderski}, {Rajarshi}, {Ramaprakash}, {Readhead}, {Reig}, {Tassis}, and
  {Zensus}]{Blinov2016}
{Blinov}, D.; {Pavlidou}, V.; {Papadakis}, I.E.; {Hovatta}, T.; {Pearson},
  T.J.; {Liodakis}, I.; {Panopoulou}, G.V.; {Angelakis}, E.; {Balokovi{\'c}},
  M.; {Das}, H.;  et~al.
\newblock {RoboPol: optical polarization-plane rotations and flaring activity
  in blazars}.
\newblock {\em \mnras} {\bf 2016}, {\em 457},~2252--2262,
  \href{http://xxx.lanl.gov/abs/1601.03392}{{\normalfont
  [arXiv:astro-ph.HE/1601.03392]}}.
\newblock {\url{https://doi.org/10.1093/mnras/stw158}}.

\bibitem[{G{\'o}mez} \em{et~al.}(2016){G{\'o}mez}, {Lobanov}, {Bruni},
  {Kovalev}, {Marscher}, {Jorstad}, {Mizuno}, {Bach}, {Sokolovsky}, {Anderson},
  {Galindo}, {Kardashev}, and {Lisakov}]{Gomez2016}
{G{\'o}mez}, J.L.; {Lobanov}, A.P.; {Bruni}, G.; {Kovalev}, Y.Y.; {Marscher},
  A.P.; {Jorstad}, S.G.; {Mizuno}, Y.; {Bach}, U.; {Sokolovsky}, K.V.;
  {Anderson}, J.M.;  et~al.
\newblock {Probing the Innermost Regions of AGN Jets and Their Magnetic Fields
  with RadioAstron. I. Imaging BL Lacertae at 21 Microarcsecond Resolution}.
\newblock {\em \apj} {\bf 2016}, {\em 817},~96,
  \href{http://xxx.lanl.gov/abs/1512.04690}{{\normalfont
  [arXiv:astro-ph.HE/1512.04690]}}.
\newblock {\url{https://doi.org/10.3847/0004-637X/817/2/96}}.

\bibitem[{Issaoun} \em{et~al.}(2022){Issaoun}, {Wielgus}, {Jorstad},
  {Krichbaum}, {Blackburn}, {Janssen}, {Chan}, {Pesce}, {G{\'o}mez}, {Akiyama},
  {Mo{\'s}cibrodzka}, {Mart{\'\i}-Vidal}, {Chael}, {Lico}, {Liu},
  {Ramakrishnan}, {Lisakov}, {Fuentes}, {Zhao}, {Moriyama}, {Broderick},
  {Tiede}, {MacDonald}, {Mizuno}, {Traianou}, {Loinard}, {Davelaar}, {Gurwell},
  {Lu}, {Alberdi}, {Alef}, {Algaba}, {Anantua}, {Asada}, {Azulay}, {Bach},
  {Baczko}, {Ball}, {Balokovi{\'c}}, {Barrett}, {Baub{\"o}ck}, {Benson},
  {Bintley}, {Blundell}, {Boland}, {Bouman}, {Bower}, {Boyce}, {Bremer},
  {Brinkerink}, {Brissenden}, {Britzen}, {Broguiere}, {Bronzwaer},
  {Bustamante}, {Byun}, {Carlstrom}, {Ceccobello}, {Chatterjee}, {Chatterjee},
  {Chen}, {Chen}, {Cho}, {Christian}, {Conroy}, {Conway}, {Cordes}, {Crawford},
  {Crew}, {Cruz-Osorio}, {Cui}, {Laurentis}, {Deane}, {Dempsey}, {Desvignes},
  {Dexter}, {Doeleman}, {Dhruv}, {Dzib Quijano}, {Eatough}, {Emami}, {Falcke},
  {Farah}, {Fish}, {Fomalont}, {Ford}, {Fraga-Encinas}, {Freeman}, {Friberg},
  {Fromm}, {Galison}, {Gammie}, {Garc{\'\i}a}, {Gentaz}, {Georgiev}, {Goddi},
  {Gold}, {G{\'o}mez-Ruiz}, {Gu}, {Hada}, {Haggard}, {Hecht}, {Hesper}, {Ho},
  {Ho}, {Honma}, {Huang}, {Huang}, {Hughes}, {Ikeda}, {Impellizzeri}, {Inoue},
  {James}, {Jannuzi}, {Jeter}, {Jiang}, {Jimenez-Rosales}, {Johnson}, {Joshi},
  {Jung}, {Karami}, {Karuppusamy}, {Kawashima}, {Keating}, {Kettenis}, {Kim},
  {Kim}, {Kim}, {Kim}, {Kino}, {Koay}, {Kocherlakota}, {Kofuji}, {Koch},
  {Koyama}, {Kramer}, {Kramer}, {Kuo}, {Bella}, {Lauer}, {Lee}, {Lee}, {Leung},
  {Levis}, {Li}, {Lico}, {Lindahl}, {Lindqvist}, {Liu}, {Liuzzo}, {Lo},
  {Lobanov}, {Lonsdale}, {Mao}, {Marchili}, {Markoff}, {Marrone}, {Marscher},
  {Matsushita}, {Matthews}, {Medeiros}, {Menten}, {Michalik}, {Mizuno},
  {Mizuno}, {Moran}, {M{\"u}ller}, {Mus}, {Musoke}, {Myserlis}, {Nadolski},
  {Nagai}, {Nagar}, {Nakamura}, {Narayan}, {Narayanan}, {Natarajan},
  {Nathanail}, {Neilsen}, {Neri}, {Ni}, {Noutsos}, {Nowak}, {Oh}, {Okino},
  {Olivares}, {Ortiz-Le{\'o}n}, {Oyama}, {{\"O}zel}, {Palumbo}, {Paraschos},
  {Park}, {Parsons}, {Patel}, {Pen}, {Pi{\'e}tu}, {Plambeck}, {PopStefanija},
  {Porth}, {P{\"o}tzl}, {Prather}, {Preciado-L{\'o}pez}, {Psaltis}, {Pu},
  {Rao}, {Rawlings}, {Raymond}, {Rezzolla}, {Ricarte}, {Ripperda}, {Roelofs},
  {Rogers}, {Ros}, {Romero-Canizales}, {Roshanineshat}, {Rottmann}, {Roy},
  {Ruiz}, {Ruszczyk}, {Rygl}, {S{\'a}nchez}, {S{\'a}nchez-Arguelles},
  {Sanchez-Portal}, {Sasada}, {Satapathy}, {Savolainen}, {Schloerb},
  {Schuster}, {Shao}, {Shen}, {Small}, {Sohn}, {SooHoo}, {Souccar}, {Sun},
  {Tazaki}, {Tetarenko}, {Tiede}, {Tilanus}, {Titus}, {Torne}, {Trent},
  {Trippe}, {van Bemmel}, {van Langevelde}, {van Rossum}, {Vos}, {Wagner},
  {Ward-Thompson}, {Wardle}, {Weintroub}, {Wex}, {Wharton}, {Wiik}, {Witzel},
  {Wondrak}, {Wong}, {Wu}, {Yamaguchi}, {Yoon}, {Young}, {Young}, {Younsi},
  {Yuan}, {Yuan}, {Zensus}, {Zhang}, and {Zhao}]{Issaoun2022}
{Issaoun}, S.; {Wielgus}, M.; {Jorstad}, S.; {Krichbaum}, T.P.; {Blackburn},
  L.; {Janssen}, M.; {Chan}, C.k.; {Pesce}, D.W.; {G{\'o}mez}, J.L.; {Akiyama},
  K.;  et~al.
\newblock {Resolving the Inner Parsec of the Blazar J1924-2914 with the Event
  Horizon Telescope}.
\newblock {\em \apj} {\bf 2022}, {\em 934},~145,
  \href{http://xxx.lanl.gov/abs/2208.01662}{{\normalfont
  [arXiv:astro-ph.HE/2208.01662]}}.
\newblock {\url{https://doi.org/10.3847/1538-4357/ac7a40}}.

\bibitem[{Zhao} \em{et~al.}(2022){Zhao}, {G{\'o}mez}, {Fuentes}, {Krichbaum},
  {Traianou}, {Lico}, {Cho}, {Ros}, {Komossa}, {Akiyama}, {Asada}, {Blackburn},
  {Britzen}, {Bruni}, {Crew}, {Dahale}, {Dey}, {Gold}, {Gopakumar}, {Issaoun},
  {Janssen}, {Jorstad}, {Kim}, {Koay}, {Kovalev}, {Koyama}, {Lobanov},
  {Loinard}, {Lu}, {Markoff}, {Marscher}, {Mart{\'\i}-Vidal}, {Mizuno}, {Park},
  {Savolainen}, and {Toscano}]{Zhao2022}
{Zhao}, G.Y.; {G{\'o}mez}, J.L.; {Fuentes}, A.; {Krichbaum}, T.P.; {Traianou},
  E.; {Lico}, R.; {Cho}, I.; {Ros}, E.; {Komossa}, S.; {Akiyama}, K.;  et~al.
\newblock {Unraveling the Innermost Jet Structure of OJ 287 with the First GMVA
  + ALMA Observations}.
\newblock {\em \apj} {\bf 2022}, {\em 932},~72,
  \href{http://xxx.lanl.gov/abs/2205.00554}{{\normalfont
  [arXiv:astro-ph.HE/2205.00554]}}.
\newblock {\url{https://doi.org/10.3847/1538-4357/ac6b9c}}.

\bibitem[{Blandford} and {Znajek}(1977)]{BZ1977}
{Blandford}, R.D.; {Znajek}, R.L.
\newblock {Electromagnetic extraction of energy from Kerr black holes.}
\newblock {\em \mnras} {\bf 1977}, {\em 179},~433--456.
\newblock {\url{https://doi.org/10.1093/mnras/179.3.433}}.

\bibitem[{Blandford} and {Payne}(1982)]{BP1982}
{Blandford}, R.D.; {Payne}, D.G.
\newblock {Hydromagnetic flows from accretion disks and the production of radio
  jets.}
\newblock {\em \mnras} {\bf 1982}, {\em 199},~883--903.
\newblock {\url{https://doi.org/10.1093/mnras/199.4.883}}.

\bibitem[Weisskopf \em{et~al.}(2022)Weisskopf, Soffitta, Baldini, Ramsey,
  O'Dell, Romani, Matt, Deininger, Baumgartner, Bellazzini, Costa,
  Kolodziejczak, Latronico, Marshall, Muleri, Bongiorno, Tennant, Bucciantini,
  Dovciak, Marin, Marscher, Poutanen, Slane, Turolla, Kalinowski, Marco,
  Fabiani, Minuti, Monaca, Pinchera, Rankin, Sgrò, Trois, Xie, Alexander,
  Allen, Amici, Andersen, Antonelli, Antoniak, Attiná, Barbanera, Bachetti,
  Baggett, Bladt, Brez, Bonino, Boree, Borotto, Breeding, Brienza, Bygott,
  Caporale, Cardelli, Carpentiero, Castellano, Castronuovo, Cavalli, Cavazzuti,
  Ceccanti, Centrone, Citraro, D'Amico, D'Alba, Gesu, Monte, Dietz, Lalla,
  Persio, Dolan, Donnarumma, Evangelista, Ferrant, Ferrazzoli, Ferrie,
  Footdale, Forsyth, Foster, Garelick, Gunji, Gurnee, Head, Hibbard, Johnson,
  Kelly, Kilaru, Lefevre, Roy, Loffredo, Lorenzi, Lucchesi, Maddox, Magazzu,
  Maldera, Manfreda, Mangraviti, Marengo, Marrocchesi, Massaro, Mauger,
  McCracken, McEachen, Mize, Mereu, Mitchell, Mitsuishi, Morbidini, Mosti,
  Nasimi, Negri, Negro, Nguyen, Nitschke, Nuti, Onizuka, Oppedisano, Orsini,
  Osborne, Pacheco, Paggi, Painter, Pavelitz, Pentz, Piazzolla, Perri,
  Pesce-Rollins, Peterson, Pilia, Profeti, Puccetti, Ranganathan, Ratheesh,
  Reedy, Root, Rubini, Ruswick, Sanchez, Sarra, Santoli, Scalise, Sciortino,
  Schroeder, Seek, Sosdian, Spandre, Speegle, Tamagawa, Tardiola, Tobia,
  Thomas, Valerie, Vimercati, Walden, Weddendorf, Wedmore, Welch, Zanetti, and
  Zanetti]{Weisskopf2022}
Weisskopf, M.C.; Soffitta, P.; Baldini, L.; Ramsey, B.D.; O'Dell, S.L.; Romani,
  R.W.; Matt, G.; Deininger, W.D.; Baumgartner, W.H.; Bellazzini, R.;  et~al.
\newblock {Imaging X-ray Polarimetry Explorer: prelaunch}.
\newblock {\em Journal of Astronomical Telescopes, Instruments, and Systems}
  {\bf 2022}, {\em 8},~1 -- 28.
\newblock {\url{https://doi.org/10.1117/1.JATIS.8.2.026002}}.

\bibitem[{Zhang} \em{et~al.}(2019){Zhang}, {Santangelo}, {Feroci}, {Xu}, {Lu},
  {Chen}, {Feng}, {Zhang}, {Brandt}, {Hernanz}, {Baldini}, {Bozzo}, {Campana},
  {De Rosa}, {Dong}, {Evangelista}, {Karas}, {Meidinger}, {Meuris}, {Nandra},
  {Pan}, {Pareschi}, {Orleanski}, {Huang}, {Schanne}, {Sironi}, {Spiga},
  {Svoboda}, {Tagliaferri}, {Tenzer}, {Vacchi}, {Zane}, {Walton}, {Wang},
  {Winter}, {Wu}, {in't Zand}, {Ahangarianabhari}, {Ambrosi}, {Ambrosino},
  {Barbera}, {Basso}, {Bayer}, {Bellazzini}, {Bellutti}, {Bertucci},
  {Bertuccio}, {Borghi}, {Cao}, {Cadoux}, {Campana}, {Ceraudo}, {Chen}, {Chen},
  {Chevenez}, {Civitani}, {Cui}, {Cui}, {Dauser}, {Del Monte}, {Di Cosimo},
  {Diebold}, {Doroshenko}, {Dovciak}, {Du}, {Ducci}, {Fan}, {Favre},
  {Fuschino}, {G{\'a}lvez}, {Gao}, {Ge}, {Gevin}, {Grassi}, {Gu}, {Gu}, {Han},
  {Hong}, {Hu}, {Ji}, {Jia}, {Jiang}, {Kennedy}, {Kreykenbohm}, {Kuvvetli},
  {Labanti}, {Latronico}, {Li}, {Li}, {Li}, {Li}, {Li}, {Limousin}, {Liu},
  {Liu}, {Lu}, {Luo}, {Macera}, {Malcovati}, {Martindale}, {Michalska}, {Meng},
  {Minuti}, {Morbidini}, {Muleri}, {Paltani}, {Perinati}, {Picciotto},
  {Piemonte}, {Qu}, {Rachevski}, {Rashevskaya}, {Rodriguez}, {Schanz}, {Shen},
  {Sheng}, {Song}, {Song}, {Sgro}, {Sun}, {Tan}, {Uttley}, {Wang}, {Wang},
  {Wang}, {Wang}, {Wang}, {Wang}, {Watts}, {Wen}, {Wilms}, {Xiong}, {Yang},
  {Yang}, {Yang}, {Yu}, {Zhang}, {Zampa}, {Zampa}, {Zdziarski}, {Zhang},
  {Zhang}, {Zhang}, {Zhang}, {Zhang}, {Zhang}, {Zhang}, {Zhang}, {Zhao},
  {Zheng}, {Zhou}, {Zorzi}, and {Zwart}]{Zhang2019}
{Zhang}, S.; {Santangelo}, A.; {Feroci}, M.; {Xu}, Y.; {Lu}, F.; {Chen}, Y.;
  {Feng}, H.; {Zhang}, S.; {Brandt}, S.; {Hernanz}, M.;  et~al.
\newblock {The enhanced X-ray Timing and Polarimetry mission{\textemdash}eXTP}.
\newblock {\em Science China Physics, Mechanics, and Astronomy} {\bf 2019},
  {\em 62},~29502,  \href{http://xxx.lanl.gov/abs/1812.04020}{{\normalfont
  [arXiv:astro-ph.IM/1812.04020]}}.
\newblock {\url{https://doi.org/10.1007/s11433-018-9309-2}}.

\bibitem[{Beechert} \em{et~al.}(2022){Beechert}, {Lazar}, {Boggs}, {Brandt},
  {Chang}, {Chu}, {Gulick}, {Kierans}, {Lowell}, {Pellegrini}, {Roberts},
  {Siegert}, {Sleator}, {Tomsick}, and {Zoglauer}]{Beechert2022}
{Beechert}, J.; {Lazar}, H.; {Boggs}, S.E.; {Brandt}, T.J.; {Chang}, Y.C.;
  {Chu}, C.Y.; {Gulick}, H.; {Kierans}, C.; {Lowell}, A.; {Pellegrini}, N.;
  et~al.
\newblock {Calibrations of the Compton Spectrometer and Imager}.
\newblock {\em Nuclear Instruments and Methods in Physics Research A} {\bf
  2022}, {\em 1031},~166510,
  \href{http://xxx.lanl.gov/abs/2203.00695}{{\normalfont
  [arXiv:astro-ph.IM/2203.00695]}}.
\newblock {\url{https://doi.org/10.1016/j.nima.2022.166510}}.

\bibitem[{Di Gesu} \em{et~al.}(2022){Di Gesu}, {Donnarumma}, {Tavecchio},
  {Agudo}, {Barnounin}, {Cibrario}, {Di Lalla}, {Di Marco}, {Escudero},
  {Errando}, {Jorstad}, {Kim}, {Kouch}, {Lindfors}, {Liodakis}, {Madejski},
  {Marshall}, {Marscher}, {Middei}, {Muleri}, {Myserlis}, {Negro}, {Omodei},
  {Pacciani}, {Paggi}, {Perri}, {Puccetti}, {Antonelli}, {Bachetti}, {Baldini},
  {Baumgartner}, {Bellazzini}, {Bianchi}, {Bongiorno}, {Bonino}, {Brez},
  {Bucciantini}, {Capitanio}, {Castellano}, {Cavazzuti}, {Ciprini}, {Costa},
  {De Rosa}, {Del Monte}, {Doroshenko}, {Dov{\v{c}}iak}, {Ehlert}, {Enoto},
  {Evangelista}, {Fabiani}, {Ferrazzoli}, {Garcia}, {Gunji}, {Hayashida},
  {Heyl}, {Iwakiri}, {Karas}, {Kitaguchi}, {Kolodziejczak}, {Krawczynski}, {La
  Monaca}, {Latronico}, {Maldera}, {Manfreda}, {Marin}, {Marinucci}, {Massaro},
  {Matt}, {Mitsuishi}, {Mizuno}, {Ng}, {O'Dell}, {Oppedisano}, {Papitto},
  {Pavlov}, {Peirson}, {Pesce-Rollins}, {Petrucci}, {Pilia}, {Possenti},
  {Poutanen}, {Ramsey}, {Rankin}, {Ratheesh}, {Romani}, {Sgr{\`o}}, {Slane},
  {Soffitta}, {Spandre}, {Tamagawa}, {Taverna}, {Tawara}, {Tennant}, {Thomas},
  {Tombesi}, {Trois}, {Tsygankov}, {Turolla}, {Vink}, {Weisskopf}, {Wu}, {Xie},
  and {Zane}]{DiGesu2022-Mrk421}
{Di Gesu}, L.; {Donnarumma}, I.; {Tavecchio}, F.; {Agudo}, I.; {Barnounin}, T.;
  {Cibrario}, N.; {Di Lalla}, N.; {Di Marco}, A.; {Escudero}, J.; {Errando},
  M.;  et~al.
\newblock {The X-ray Polarization View of
  Mrk\raisebox{-0.5ex}\textasciitilde421 in an Average Flux State as Observed
  by the Imaging X-ray Polarimetry Explorer}.
\newblock {\em arXiv e-prints} {\bf 2022}, p. arXiv:2209.07184,
  \href{http://xxx.lanl.gov/abs/2209.07184}{{\normalfont
  [arXiv:astro-ph.HE/2209.07184]}}.

\end{thebibliography}

\end{adjustwidth}

\end{document}